\documentclass[aps,12pt,showpacs,preprint,groupedaddress,floatfix]{revtex4}
\usepackage{graphicx}
\usepackage{dcolumn}
\usepackage{bm}
\usepackage{amssymb}
\usepackage{amsmath}
\usepackage{color}

\newcommand{\bff}[1]{{\mbox{\boldmath $#1$}}}

\newcommand{\bra}[1]{\left\langle #1 \right|}
\newcommand{\ket}[1]{\left| #1 \right\rangle}

\begin{document}

\title{3D Relativistic Hartree-Bogoliubov model with a separable pairing interaction}
\author{T. Nik\v si\' c}
\affiliation{Physics Department, Faculty of Science, University of Zagreb,
10000 Zagreb, Croatia}
\author{D. Vretenar}
\affiliation{Physics Department, Faculty of Science, University of Zagreb,
10000 Zagreb, Croatia}
\author{Yuan Tian}
\affiliation{China Institute of Atomic Energy, Beijing 102413, People's Republic of
   China}
\author{Zhong-yu Ma}
\affiliation{China Institute of Atomic Energy, Beijing 102413, People's Republic of
   China}
\author{P. Ring}
\affiliation{Physik-Department der Technischen Universit\"at M\"unchen,
D-85748  Garching, Germany}
\date{\today}

\begin{abstract}
A recently introduced separable pairing force for relativistic
Hartree-Bogoliubov (RHB) calculations, adjusted in nuclear matter to
the pairing gap of the Gogny force, is employed in the 3D RHB model
for triaxial shapes. The pairing force is separable in momentum space
but, when transformed to coordinate space in calculations of finite
nuclei, it is no longer separable because of translational
invariance. The corresponding pairing matrix elements are represented as
a sum of a finite number of separable terms in the basis of a 3D
harmonic oscillator. The 3D RHB model is applied to the calculation
of binding energy surfaces and pairing energy maps  
for a sequence of even-A Sm isotopes.
\end{abstract}

\maketitle

\section{\label{secI}Introduction}
The structure of heavy complex nuclei with a large number of active
valence nucleons is, at present, best described by the framework of
nuclear energy density functionals (NEDF). A variety of structure
phenomena, not only in stable nuclei, but also in regions of nuclei
far from the valley of $\beta$-stability and close to the nucleon
drip-lines, have been described with self-consistent mean-field
models based on the Gogny interaction, the Skyrme energy functional,
and the relativistic meson-exchange effective
Lagrangians~\cite{BHR.03,VALR.05}.

Self-consistent relativistic mean-field (RMF) models have been
employed in analyses of properties of ground and excited states in
spherical and deformed nuclei. For a quantitative analysis of
open-shell nuclei it is necessary to consider also pairing
correlations. Pairing has often been taken into account in a very
phenomenological way in the BCS model with the monopole pairing
force, adjusted to the experimental odd-even mass differences. 
In many cases, however, this approach presents only a poor 
approximation. The physics of weakly-bound nuclei, in particular,
necessitates a unified and self-consistent treatment of mean-field
and pairing correlations. This has led to the formulation and
development of the relativistic Hartree-Bogoliubov (RHB) model, which
represents a relativistic extension of the conventional
Hartree-Fock-Bogoliubov framework. In most applications of the RHB
model~\cite{VALR.05} the pairing part of the Gogny
force~\cite{BGG.84} has be employed in the particle-particle ($pp$)
channel:
\begin{equation}
V^{pp}(1,2)~=~\sum_{i=1,2}e^{-((\mathbf{r}_{1}-\mathbf{r}_{2})/{\mu_{i}})^{2}%
}\,(W _{i}~+~B_{i}P^{\sigma}-H_{i}P^{\tau}-M_{i}P^{\sigma}P^{\tau})\;,
\label{Gogny}
\end{equation}
with the set D1S~\cite{BGG.91} for the parameters $\mu_{i}$, $W_{i}$,
$B_{i}$, $H_{i}$, and $M_{i}$ $(i=1,2)$. A basic advantage of the
Gogny force is the finite range, which automatically guarantees a
proper cut-off in momentum space. However, the resulting pairing
field is non-local and the solution of the corresponding
Dirac-Hartree-Bogoliubov integro-differential equations can be
time-consuming, especially in the case of deformed nuclei. An
alternative is the use of a zero-range, possibly density-dependent,
$\delta$-force in the $pp$-channel of the RHB model~\cite{Meng.98},
but this approach introduces an additional cut-off parameter in
energy. The effective range of the pairing interaction is determined
by the energy cut-off, and the strength parameter must be chosen
accordingly in order to reproduce empirical pairing gaps. In
Ref.~\cite{NRV.05} we have  implemented a renormalization scheme for
the relativistic Hartree-Bogoliubov equations with a zero-range
pairing interaction. The procedure is equivalent to a simple energy
cut-off with a position dependent coupling constant, and the
resulting average pairing gaps and pairing energies do not depend on
the cut-off energy. A density-dependent strength parameter of the
zero-range pairing can be adjusted in such a way that the
renormalization procedure reproduces in symmetric nuclear matter the
pairing gap of the Gogny force Eq.~(\ref{Gogny}).

In a series of recent articles~\cite{TMR.09a,TMR.09b,TMR.09c} we have
introduced a separable form of the pairing force for RHB calculations
in spherical and axially deformed nuclei. The force is separable in
momentum space, and is completely determined by two parameters that
are adjusted to reproduce in symmetric nuclear matter the bell-shape
curve of the pairing gap of the Gogny force. Technically this approach is 
similar to the derivation of non-empirical pairing functionals 
from low-momentum interactions \cite{DG.08,LDBM.09}. In applications to
finite nuclei, when transformed from momentum to coordinate space,
the pairing force is no longer separable because of translational
invariance. It has been shown, however, that a method developed by
Talmi and Moshinsky can be used to represent the corresponding $pp$
matrix elements as a sum of a finite number of separable terms. When
the nucleon wave functions are expanded in a harmonic oscillator
basis, spherical or axially deformed, the sum converges relatively
quickly, i.e. a reasonably small number of separable terms reproduces
with high accuracy the results of calculations performed in a
complete basis. 

The simple separable force considered in
Refs.~\cite{TMR.09a,TMR.09b,TMR.09c} reproduces pairing properties of
spherical and axially deformed nuclei calculated with the original
Gogny force, but  with the important advantage that the computational
cost is greatly reduced. In the present work we extend this approach
to triaxial nuclei, and introduce a 3D RHB model with the separable
pairing interaction of Ref.~\cite{TMR.09a} in the $pp$ channel. This
model will enable systematic calculations of RHB binding energy
surfaces in the $\beta-\gamma$ plane, based on relativistic nuclear
energy density functionals, that can be used as input for the
generator coordinate method configuration mixing of angular-momentum
projected triaxial wave functions, or to determine the parameters of
a five-dimensional collective Hamiltonian for quadrupole vibrational
and rotational degrees of freedom~\cite{Ni.09}. In Sec.~\ref{secII}
we introduce the 3D RHB model and derive the $pp$ matrix element of
the pairing force as a sum of a finite number of separable terms in
the basis of a 3D harmonic oscillator. Illustrative calculations for
even-A Sm isotopes are presented in In Sec.~\ref{secIII}.
Section \ref{conclusions} summarizes the results and ends with 
an outlook for future applications. Details on the
expansion of single-nucleon spinors in the 3D harmonic oscillator
basis, and the transformation of the product of harmonic oscillator
wave functions to the center-of-mass frame, are included in the two
appendixes.

\section{\label{secII} 3D Relativistic Hartree-Bogoliubov model
with a separable pairing interaction}

The relativistic Hartree-Bogoliubov framework~\cite{VALR.05} provides
a unified description of particle-hole $(ph)$ and particle-particle
$(pp)$ correlations on a mean-field level by using two average
potentials: the self-consistent mean field that encloses all the long
range \textit{ph} correlations, and a pairing field $\hat{\Delta}$
which sums up the \textit{pp}-correlations. The ground state of a
nucleus is described by a generalized Slater determinant
$|\Phi\rangle$ that represents the vacuum with respect to independent
quasiparticles. The quasiparticle operators are defined by the
unitary Bogoliubov transformation of the single-nucleon creation and
annihilation operators:
\begin{equation}
\alpha_{k}^{+}=\sum\limits_{l}U_{lk}^{{}}c_{l}^{+}+V_{lk}^{{}}c_{l}^{{}}\;,
\end{equation}
where $U$ and $V$ are the Hartree-Bogoliubov wave functions determined by
the solution of the RHB equation. In coordinate representation:
\begin{equation}
\label{eq:RHB}
\left( \begin{array}{cc}
   h_D-m-\lambda  &  \Delta \\
   -\Delta^*  & -h_D^*+m+\lambda
\end{array} \right) \left( \begin{array}{c} U_k(\bff{r}) \\ V_k(\bff{r}) \end{array} \right)
= E_k    \left( \begin{array}{c} U_k(\bff{r}) \\ V_k(\bff{r}) \end{array} \right) \;.
\end{equation}
In the relativistic case the self-consistent mean-field corresponds
to the single-nucleon Dirac Hamiltonian $\hat{h}_D$. In the usual
$\sigma$, $\omega$, and $\rho$ meson-exchange representation, and for
the stationary case with time-reversal symmetry, i.e. for the
ground-state of an even-even nucleus:
\begin{equation}
\hat{h}_D =  -i\bff{\alpha} \bff{\nabla} + \beta(m+g_\sigma \sigma(\bff{r}))
   +g_\omega \omega^0(\bff{r}) + g_\rho \tau_3 \rho^0(\bff{r})+
   e\frac{1-\tau_3}{2}A^0(\bff{r}) \; .
\end{equation}
The classical meson fields are solutions of the stationary
Klein-Gordon equations:
\begin{align}
\left[  -\triangle+m_{\sigma}^{2}\right]  \sigma(\bm{r})  &  =-g_{\sigma}%
(\rho)\rho_{s}(\bm{r})\;,\label{KG-sigma}\\
\left[  -\triangle+m_{\omega}^{2}\right]  \omega^{0}(\bm{r})  &  =g_{\omega
}(\rho)\rho(\bm{r})\;,\label{KG-omega}\\
\left[  -\triangle+m_{\rho}^{2}\right]  {\rho}~^{0}(\bm{r})  &
=g_{\rho}(\rho)\rho_3^{0}(\bm{r})\;,\label{KG-rho}\\
-\triangle A^{0}(\bm{r})  &  =\rho_{p}(\bm{r})\;, \label{Poisson_4}%
\end{align}
for the $\sigma$ meson, the time-like components of the $\omega$
meson and  $\rho$ meson, and the Poisson equation for the vector
potential, respectively. In the general case when the meson-nucleon
couplings $g_{\sigma}$, $g_{\omega}$, and $g_{\rho}$ explicitly
depend on the nucleon (vector) density $\rho$, there is an additional
contribution to the nucleon self-energy - the rearrangement
term \cite{VALR.05}, essential for the energy-momentum conservation and the
thermodynamical consistency of the model.

In Eq.~(\ref{eq:RHB}) $m$ is the nucleon mass, and
the chemical potential $\lambda$ is determined by the particle
number subsidiary condition in order that the expectation value of
the particle number operator in the ground state equals the number
of nucleons. The pairing field $\Delta$ reads
\begin{equation}
\Delta_{ab}(\bff{r},\bff{r}^\prime) = \frac{1}{2}\sum_{c,d}{V_{abcd}(\bff{r},\bff{r}^\prime)}
\kappa_{cd}(\bff{r},\bff{r}^\prime).
\end{equation}
where $V_{abcd}(\bff{r},\bff{r}^\prime)$ are the matrix elements of
the two-body pairing interaction, and the indices $a$, $b$, $c$ and
$d$ denote the quantum numbers that specify the Dirac indices of the
spinor. The column vectors denote the quasiparticle wave functions,
and $E_{k}$ are the quasiparticle energies. The dimension of the RHB
matrix equation is two times the dimension of the corresponding Dirac
equation. For each eigenvector $(U_{k} ,V_{k} )$ with positive
quasiparticle energy $E_{k} > 0$, there exists an eigenvector
$(V_{k}^{*},U_{k}^{*})$ with quasiparticle energy $-E_{k}$. Since the
baryon quasiparticle operators satisfy fermion commutation relations,
the levels $E_{k}$ and $-E_{k}$ cannot be occupied simultaneously.
For the solution that corresponds to a ground state of a nucleus with
even particle number, one usually chooses the eigenvectors with
positive eigenvalues $E_{k}$.

The single-particle density and the pairing tensor, constructed from
the quasi-particle wave functions
\begin{align}
\label{eq:single-particle-density}
\rho_{cd}(\bff{r},\bff{r}^\prime) &= \sum_{k>0}{V^*_{ck}(\bff{r})V_{dk}(\bff{r}^\prime)},\\
\label{eq:pairing-tensor}
\kappa_{cd}(\bff{r},\bff{r}^\prime)&=\sum_{k>0}{U^*_{ck}(\bff{r})V_{dk}(\bff{r}^\prime)},
\end{align}
are calculated in the \emph{no-sea} approximation (denoted by $k>0$):
the summation runs over all quasiparticle states $k$
with positive quasiparticle energies $E_k>0$, but omits states that
originate from the Dirac sea. The latter are characterized by a
quasiparticle energy larger than the Dirac gap ($\approx 1200$ MeV).

Pairing correlations in nuclei are restricted to an energy window of
a few MeV around the Fermi level, and their scale is well separated
from the scale of binding energies, which are in the range of several
hundred to thousand MeV. There is no empirical evidence for any
relativistic effect in the nuclear pairing field $\hat{\Delta}$ and,
therefore, a hybrid RHB model with a non-relativistic pairing
interaction can be formulated. For a general two-body interaction,
the matrix elements of the relativistic pairing field read
\begin{equation}
\hat{\Delta}_{a_1 p_1, a_2 p_2} =
{\frac{1}{2}}\sum\limits_{a_3 p_3, a_4 p_4}
\langle a_1 p_1, a_2 p_2 |V^{pp}|a_3 p_3, a_4 p_4\rangle_a~
\kappa_{a_3 p_3, a_4 p_4}\; ,
\end{equation}
where the indices ($p_1,p_2,p_3,p_4 \equiv f, g$) refer to the large
and small components of the quasiparticle Dirac spinors:
\begin{equation}
U({\bf r},s,t)\ =\
\left( \begin{array}{c}
f_{U}({\bf r},s,t) \\ ig_{U}({\bf r},s,t)
\end{array} \right)  \quad\quad\quad
V({\bf r},s,t)\ =\
\left( \begin{array}{c}
f_{V}({\bf r},s,t) \\ ig_{V}({\bf r},s,t)
\end{array} \right) \; .
\label{UV}
\end{equation}
In practical applications of the RHB model to finite open-shell
nuclei, only the large components of the spinors $U_{k}({\bf r})$ and
$V_{k}({\bf r})$ are used to build the non-relativistic pairing
tensor $\hat{\kappa}$ in Eq.~(\ref{eq:pairing-tensor}). The resulting
pairing field reads
\begin{equation}
\hat{\Delta}_{a_1 f, a_2 f} =
{\frac{1}{2}}\sum\limits_{a_3 f, a_4 f}
\langle a_1 f, a_2 f |V^{pp}|a_3 f, a_4 f\rangle_a~
\kappa_{a_3 f, a_4 f}\; .
\end{equation}
The other components: $\hat{\Delta}_{fg}$, $\hat{\Delta}_{gf}$, and
$\hat{\Delta}_{gg}$ can be safely omitted~\cite{SR.02}.

The Dirac-Hartree-Bogoliubov equations and the equations for the
meson fields are solved by expanding the nucleon spinors $U({\bf r},
s, t)$ and $V({\bf r}, s, t)$, and the meson fields, in the basis of
a three-dimensional harmonic oscillator (HO) in Cartesian
coordinates. In this way both axial and triaxial nuclear shapes can
be described. In addition, to reduce the computational task, it is
assumed that  the total densities are symmetric under reflections
with respect to all three planes $xy$, $xz$ and $yz$. When combined
with time-reversal invariance, this also implies that parity is
conserved. The single-nucleon basis is defined in
Appendix~\ref{App-basis}.

In Ref.~\cite{TMR.09a} a new separable form of the pairing
interaction has been introduced, with parameters adjusted to
reproduce the pairing properties of the Gogny force in nuclear
matter. The gap equation in the $^1$S$_0$ channel reads
\begin{equation}
\Delta(k) = - \int_0^{\infty} {{k'^2 dk'}\over{2\pi^2}} \bra{k}V^{^1S_0}\ket{k'}
{{\Delta(k')} \over {2E(k')}} \; ,
\end{equation}
and the pairing force is separable in momentum space:
\begin{equation}
\label{sep_pair}
\bra{k}V^{^1S_0}\ket{k'} = - G p(k) p(k') \;.
\end{equation}
By assuming a simple Gaussian ansatz $p(k) = e^{-a^2k^2}$, the two
parameters $G$ and $a$ have been adjusted to reproduce the density
dependence of the gap at the Fermi surface, calculated with a Gogny
force. For the D1S parameterization~\cite{BGG.91} of the Gogny force:
$G=-728\;{\rm MeV fm}^3$ and $a=0.644\;{\rm fm}$. When the pairing
force Eq.~(\ref{sep_pair}) is transformed from momentum to coordinate
space, it takes the form:
\begin{equation}
\label{pp-force}
V(\bff{r}_1,\bff{r}_2,\bff{r}_1^\prime,\bff{r}_2^\prime)
=G\delta \left(\bff{R}-\bff{R}^\prime \right)P(\bff{r})P(\bff{r}^\prime)
\frac{1}{2}\left(1-P^\sigma \right),
\end{equation}
where $\bff{R}=\frac{1}{2}\left(\bff{r}_1+\bff{r}_2\right)$ and
$\bff{r}=\bff{r}_1-\bff{r}_2$ denote the center-of-mass and the relative coordinates,
and $P(\bff{r})$ is the Fourier transform of $p(k)$:
\begin{equation}
P(\bff{r}) = \frac{1}{\left(4\pi a^2 \right)^{3/2}}e^{-\bff{r}^2/4a^2} \;.
\end{equation}
The pairing force has finite range and, because of the presence of
the factor $\delta\left(\bff{R}-\bff{R}^\prime \right)$, it preserves
translational invariance. Even though
$\delta\left(\bff{R}-\bff{R}^\prime \right)$ implies that this force
is not completely separable in coordinate space, the corresponding
$pp$ matrix elements can be represented as a sum of a finite number
of separable terms in the basis of a 3D harmonic oscillator:
\begin{equation}
\label{eq:matrix-element}
V_{\alpha \beta \gamma \delta}^{pp} =
\sum_{N_x=0}^{N_x^0}\sum_{N_y=0}^{N_y^0}\sum_{N_z=0}^{N_z^0}
V_{\alpha \beta}^{N_xN_yN_z}V_{\gamma \delta}^{N_xN_yN_z} \; ,
\end{equation}
where $N_x$, $N_y$, and $N_z$ are the quantum numbers of the
corresponding 1D HOs in the center-of-mass frame (cf.
Appendix~\ref{App-pp}). This means that the pairing field can also be
written as a sum of a finite number of separable terms:
\begin{equation}
\label{eq:pairing-field}
\Delta_{\alpha \beta} = \frac{1}{2}\sum_{N_x=0}^{N_x^0}\sum_{N_y=0}^{N_y^0}
     \sum_{N_z=0}^{N_z^0}{P_{N_xN_yN_z} V_{\alpha \beta}^{N_xN_yN_z}} \; ,
\end{equation}
with the coefficients
\begin{equation}
\label{eq:separable-parameters}
P_{N_xN_yN_z} = \sum_{\gamma \delta}{V_{\gamma \delta}^{N_xN_yN_z}
        \kappa_{\gamma \delta}} \; .
\end{equation}
The advantage of using the separable pairing interaction
Eq.~(\ref{pp-force}) is that the matrices $V_{\alpha
\beta}^{N_xN_yN_z}$ are calculated only once at the beginning of a
self-consistent calculation. The coefficients $P_{N_xN_yN_z}$ are
re-calculated at each iteration step, using the corresponding updated
pairing tensor $\kappa$.

In the following we calculate the antisymmetric matrix element  of
the pairing interaction Eq.~(\ref{pp-force})
\begin{equation}
\bra{\alpha \bar{\gamma}} V \ket{\beta \bar{\delta}}_a =
\bra{\alpha \bar{\gamma}} V \ket{\beta \bar{\delta}}
-\bra{\alpha \bar{\gamma}} V \ket{\bar{\delta}\beta} \;,
\end{equation}
in the basis of a 3D harmonic oscillator, where $\ket{\alpha}$ and
$\ket{\bar \alpha}$ denote the positive and negative x-simplex
operator eigenstates (cf. Appendix~\ref{App-basis})
\begin{align}
\ket{\alpha} \equiv \ket{n_x^\alpha n_y^\alpha n_z^\alpha;\;i=+}
   &= \ket{n^\alpha}
   \frac{i^{n_y^\alpha}}{\sqrt{2}}\left[\ket{\uparrow} -(-1)^{n_x^\alpha}\ket{\downarrow} \right],\\
\ket{\bar \alpha} \equiv \ket{n_x^\alpha n_y^\alpha n_z^\alpha ;\;i=-}
   &= \ket{n^\alpha}\frac{i^{n_y^\alpha}}{\sqrt{2}}(-1)^{n_x^\alpha+n_y^\alpha+1}
   \left[\ket{\uparrow} +(-1)^{n_x^\alpha}\ket{\downarrow} \right] \;.
\end{align}
The matrix element can be separated into a product of spin and
coordinate space factors
\begin{equation}
\bra{\alpha \bar{\gamma}} V \ket{\beta \bar{\delta}}_a =
\bra{\alpha \bar{\gamma}}W\frac{1}{2}(1-P^\sigma)\ket{\beta \bar{\delta}}_a \;.
\end{equation}
The operator  $\frac{1}{2}(1-P^\sigma)$ projects onto the $S=0$
spin-singlet product state
\begin{equation}
\ket{\beta \bar{\delta}}_{S=0}=-\ket{ \bar{\delta}\beta}_{S=0}
=\frac{1}{4}i^{n_y^\delta+n_y^\beta}\left(-1\right)^{n_y^\delta+1}
\left[ 1 + (-1)^{n_x^\beta+n_x^\delta} \right] \left[ \ket{\uparrow\downarrow}
-\ket{\downarrow\uparrow} \right] \ket{n^\beta n^\delta } \; ,
\end{equation}
and the problem is reduced to the calculation of the spatial part of
the matrix element
\begin{align}
\bra{\alpha \bar{\gamma}} V \ket{\beta \bar{\delta}}_a &=
\frac{1}{8}i^{n_y^\alpha+n_y^\beta+n_y^\gamma+n_y^\delta}
(-1)^{n_y^\alpha+n_y^\delta}\left[1+(-1)^{n_x^\beta+n_x^\delta} \right]
\left[ 1+(-1)^{n_x^\alpha+n_x^\gamma}\right] \times \\
&\times \left[ \bra{n^\alpha n^\gamma} W \ket{n^\beta n^\delta}
     + \bra{n^\alpha n^\gamma} W \ket{n^\delta n^\beta }  \right] \; .
\end{align}
For $W(\bff{r}_1,\bff{r}_2,\bff{r}_1^\prime,\bff{r}_2^\prime)
= G\delta\left(\bff{R}-\bff{R}^\prime\right)P(\bff{r})P(\bff{r}^\prime)$
(cf. Eq.~(\ref{pp-force}) ),
the spatial part of the matrix element
\begin{equation}
I \equiv \int{\phi_{n_\alpha}(\bff{r}_1) \phi_{n_\gamma}(\bff{r}_2)
W(\bff{r}_1,\bff{r}_2,\bff{r}_1^\prime,\bff{r}_2^\prime)
  \phi_{n_\beta}(\bff{r}_1^\prime) \phi_{n_\delta}(\bff{r}_2^\prime)}
  d\bff{r}_1d\bff{r}_2d\bff{r}_1^\prime d\bff{r}_2^\prime \; ,
\end{equation}
can be decomposed into three Cartesian components
\begin{equation}
\label{I-tot}
I = G I_x I_y I_z \; .
\end{equation}
Here we only derive a detailed expression for the $x$-component
\begin{equation}
\label{integral-x}
I_x = \int \phi_{n_x^\alpha}(x_1,b_x)\phi_{n_x^\gamma}(x_2,b_x)
   P(x) \delta(X-X^\prime) P(x^\prime)
   \phi_{n_x^\beta}(x_1^\prime,b_x)\phi_{n_x^\delta}(x_2^\prime,b_x)
   dx_1dx_2dx_1^\prime dx_2^\prime \; .
\end{equation}
We will use the transformation of a product of 1D HO wave functions
to the center-of-mass and relative coordinates, derived in
Eq.~(\ref{moshinsky})
\begin{align}
\phi_{n_x^\alpha}(x_1,b_x)\phi_{n_x^\gamma}(x_2,b_x) &=
   \sum_{N_x,n_x}{M_{n_x^\alpha n_x^\gamma}^{n_xN_x}
             \phi_{N_x}(X,B)\phi_{n_x}(x,b_{x,r})}, \\
\phi_{n_x^\beta}(x_1^\prime,b_x)\phi_{n_x^\delta}(x_2^\prime,b_x) &=
   \sum_{N_x^\prime,n_x^\prime}{M_{n_x^\beta n_x^\delta}^{n_x^\prime N_x^\prime}
             \phi_{N^\prime_x}(X^\prime,B)\phi_{n^\prime_x}(x^\prime,b_{x,r})} \; .
\end{align}
The volume element in Eq.~(\ref{integral-x}) is transformed:
\begin{equation}
dx_1dx_2dx_1^\prime dx_2^\prime
=-dXdX^\prime dx dx^\prime.
\end{equation}
The integral Eq.~(\ref{integral-x}) now reads
\begin{align}
I_x &= -\sum_{N_x,n_x}M_{n_\alpha^x n_\gamma^x}^{n_xN_x}
   \sum_{N^\prime_x,n^\prime_x}M_{n_\beta^x n_\delta^x}^{n_x^\prime N_x^\prime}
   \int{P(x)\phi_{n_x}(x,b_{x,r})dx}
\int{P(x^\prime)\phi_{n_x^\prime}(x^\prime,b_{x,r})dx^\prime}\times\nonumber \\
 &\times \iint{\delta(X-X^\prime)\phi_{N_x}(X,B)  \phi_{N^\prime_x}(X^\prime,B)
   dXdX^\prime}.
\end{align}
By making use of:
\begin{equation}
\iint{\delta(X-X^\prime)\phi_{N_x}(X,B)  \phi_{N^\prime_x}(X^\prime,B)dXdX^\prime }
=\int{\phi_{N_x}(X,B)  \phi_{N^\prime_x}(X,B)dX} =\delta_{N_x,N_x^\prime}\; ,
\end{equation}
the integral can be written in the form
\begin{equation}
\label{Ix}
I_x = -\sum_{N_x}M_{n_\alpha^x n_\gamma^x}^{n_xN_x}
M_{n_\beta^x n_\delta^x}^{n_x^\prime N_x}
I_{n_x}(b_{x,r})I_{n_x^\prime}(b_{x,r}) \;,
\end{equation}
where $I_{n_x}$ denotes
\begin{equation}
\label{In}
I_{n_x}(b_{x,r}) =  \int{P(x)\phi_{n_x}(x,b_{x,r})dx} \;.
\end{equation}
Note that the conditions
\begin{equation}
n^x_\alpha+n^x_\gamma = n_x + N_x \quad \textnormal{and} \quad
n^x_\beta+n^x_\delta = n_x^\prime + N_x^\prime \; ,
\end{equation}
have been used to eliminate the sums over $n_x$ and $n_x^\prime$. To
evaluate the integral $I_{n_x}(b_{x,r})$,  we make use of the
generating function for the HO wave functions
Eq.~(\ref{generating-HO}), and calculate the following integral:
\begin{align}
\label{JB1}
J(p,b) &= \int_{-\infty}^{\infty}{g(x,p,b)P(x)dx}
 =\pi^{-1/4} \sqrt{\frac{b}{b^2+2a^2}}
 \sum_{n=0}^\infty{(-1)^n\left(\frac{b^2-2a^2}{b^2+2a^2}  \right)^np^{2n}} \; .
\end{align}
Using the definition of the generating function
Eq.~(\ref{generating-HO}), this expression can also be written as
follows
\begin{equation}
\label{JB2}
J(p,b) = \sum_{n=0}^\infty{p^{2n}
        \sqrt{\frac{2^{2n}}{(2n)!}} \int_{-\infty}^\infty{P(x)\phi_{2n}(x,b)} } \; .
\end{equation}
The series contains only even powers because $P(x)$ is a symmetric
function. By comparing Eqs.~(\ref{JB1}) and (\ref{JB2}), we obtain
\begin{equation}
 \int_{-\infty}^\infty{P(x)\phi_{n}(x,b)}
= \pi^{-1/4} \sqrt{\frac{b}{b^2+2a^2}}
 (-1)^{n/2}\left(\frac{b^2-2a^2}{b^2+2a^2}  \right)^{n/2}\delta_{n,even}\; ,
\end{equation}
and finally, inserting the relative oscillator length $b_r = \sqrt{2}b$
(cf. Appendix~\ref{App-pp}) ,
\begin{equation}
I_{n_x}(b_{x,r})
= \frac{1}{(2\pi)^{1/4}} \sqrt{\frac{b}{b^2+a^2}}
 (-1)^{n/2}\left(\frac{b^2-a^2}{b^2+a^2}  \right)^{n/2}\delta_{n,even} \; .
\end{equation}

The number of terms in Eq.~(\ref{Ix}) and then, of course, in
Eqs.~(\ref{eq:matrix-element}) and (\ref{eq:pairing-field}) is in
principle limited by the dimension of the oscillator basis. If
single-particle oscillator states $\ket{n_xn_yn_z}$ with $n_x+n_y+n_z
\le N_f^{max}$ form the basis, the summation over the quantum number
of the 1D HO in the center-of-mass frame in Eq.~(\ref{Ix}) runs over
$N_x=0,\dots,2N_f^{max}$. This means that the maximal total number of
terms in Eqs.~(\ref{eq:matrix-element}) and (\ref{eq:pairing-field})
equals $N_{tot}^{max}=(2N_f^{max}+1)^3$. However, results of
calculations performed in Refs.~\cite{TMR.09a,TMR.09c} suggest that
the actual number of terms that give significant contributions to the
pairing field is much smaller. If a cut-off condition is imposed
\begin{equation}
N_x \le N_x^c,\quad N_y \le N_y^c, \quad N_z \le N_z^c \;,
\end{equation}
the total number of separable terms becomes:
\begin{equation}
N_{\rm sep}=\frac{1}{8}\left(N_x^c+2\right)\left(N_y^c+2\right)\left(N_z^c+2\right) \;.
\end{equation}
In the next section we will compare some results of illustrative 3D
RHB calculations with those obtained assuming axial
symmetry~\cite{TMR.09c}. For a meaningful comparison with results
calculated using the axial RHBZ code~\cite{TMR.09c}, we will make the
following choice: $N_x+N_y \le N_\perp^c$ and $N_z\le N_z^c$. In this
case the total number of separable terms equals
\begin{equation}
N^{\rm axial}_{\rm sep}=\frac{1}{8}\left(N_\perp^c+1  \right)^2\left(N_z^c+2  \right).
\end{equation}

\section{\label{secIII}Illustrative calculations}

In this section we present the results of illustrative 3D RHB
calculations for a sequence of Sm isotopes. The separable pairing
force Eq.~(\ref{pp-force}) is used in the $pp$ channel, and the mean
field is determined by the density-dependent meson-exchange effective
interaction DD-ME2~\cite{LNVR.05} in the $ph$ channel. DD-ME2 has
been adjusted to empirical properties of symmetric and asymmetric
nuclear matter, binding energies, charge radii, and neutron radii of
spherical nuclei. The interaction has been tested in calculations of
ground state properties of large set of spherical and deformed
nuclei. An excellent agreement with data has been obtained for
binding energies, charge isotope shifts, and quadrupole deformations.
When used in the relativistic RPA, DD-ME2 reproduces with high
accuracy data on isoscalar and isovector collective
excitations~\cite{LNVR.05,PVKC.07}.

\begin{figure}
\begin{tabular}{cc}
\includegraphics[scale=0.45]{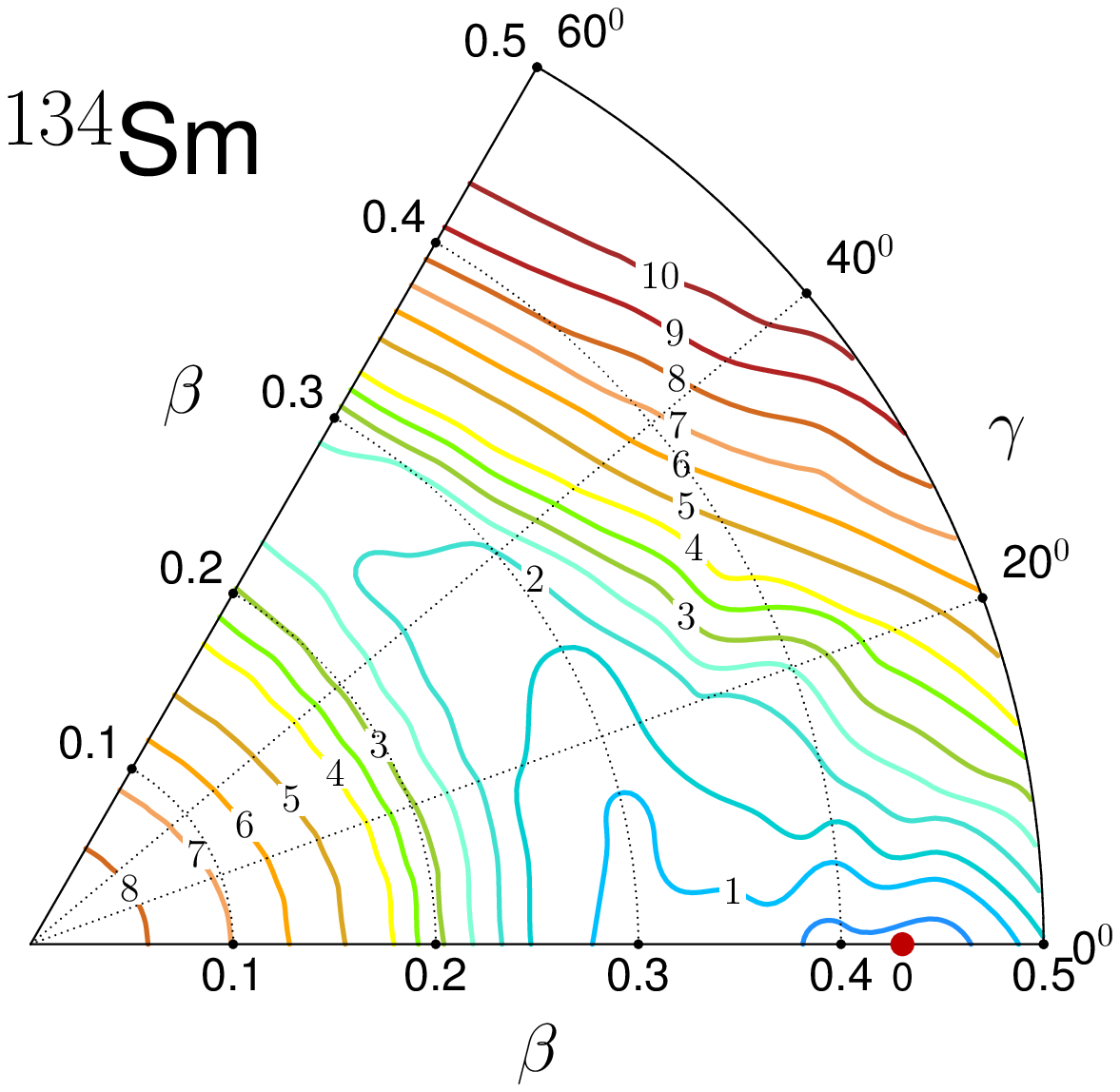}&
\includegraphics[scale=0.45]{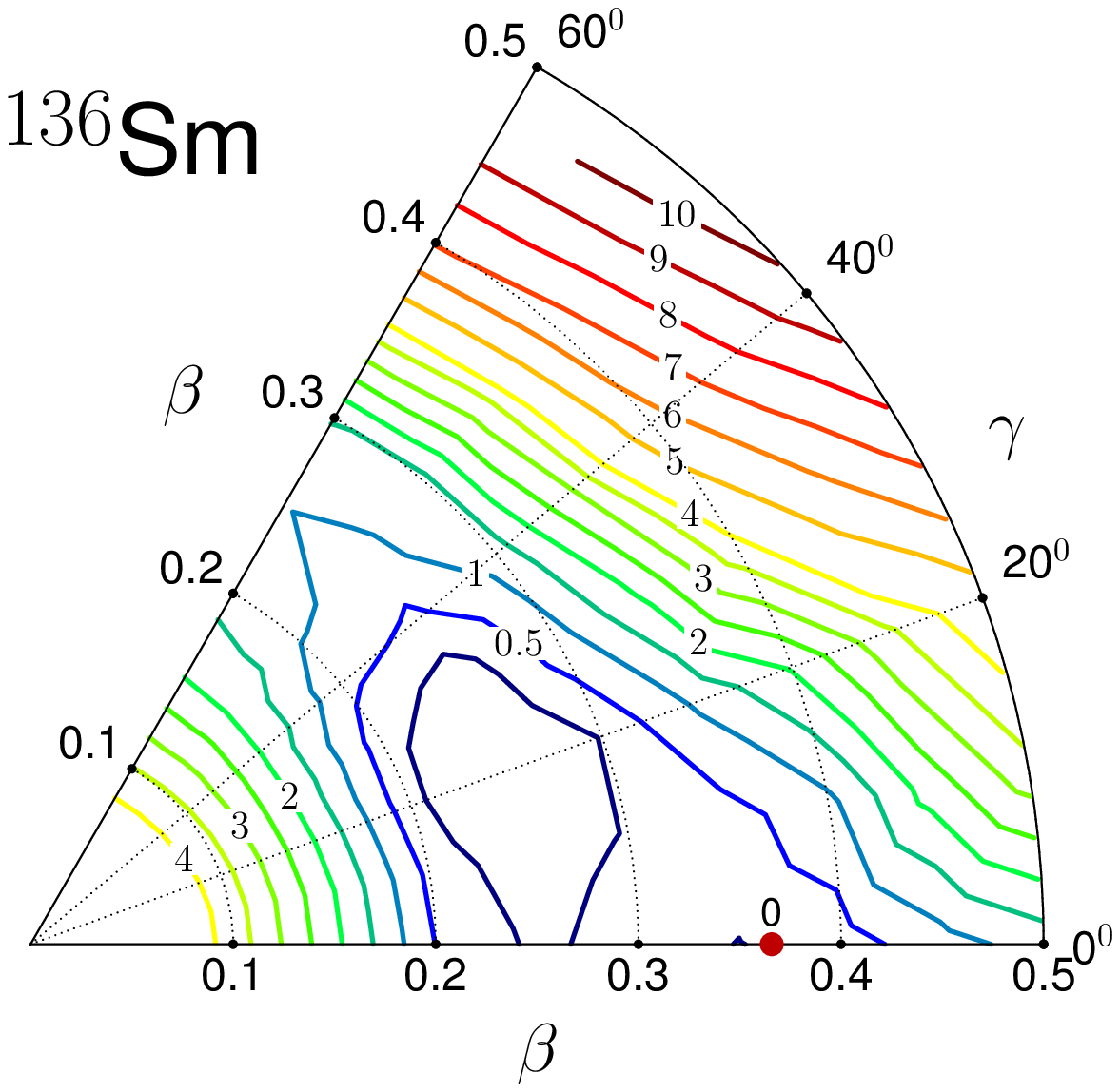}\\
\includegraphics[scale=0.45]{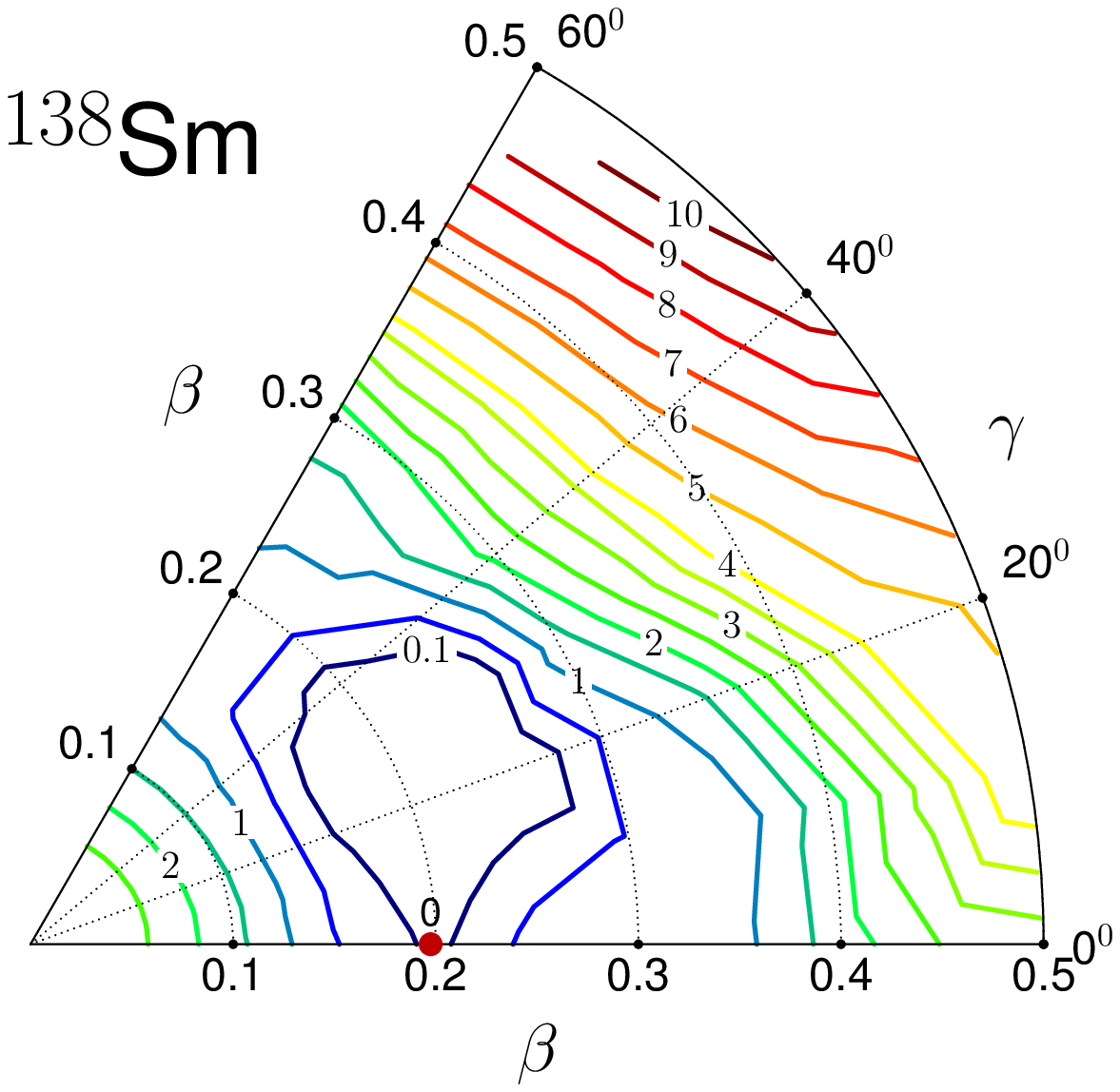} &
\includegraphics[scale=0.45]{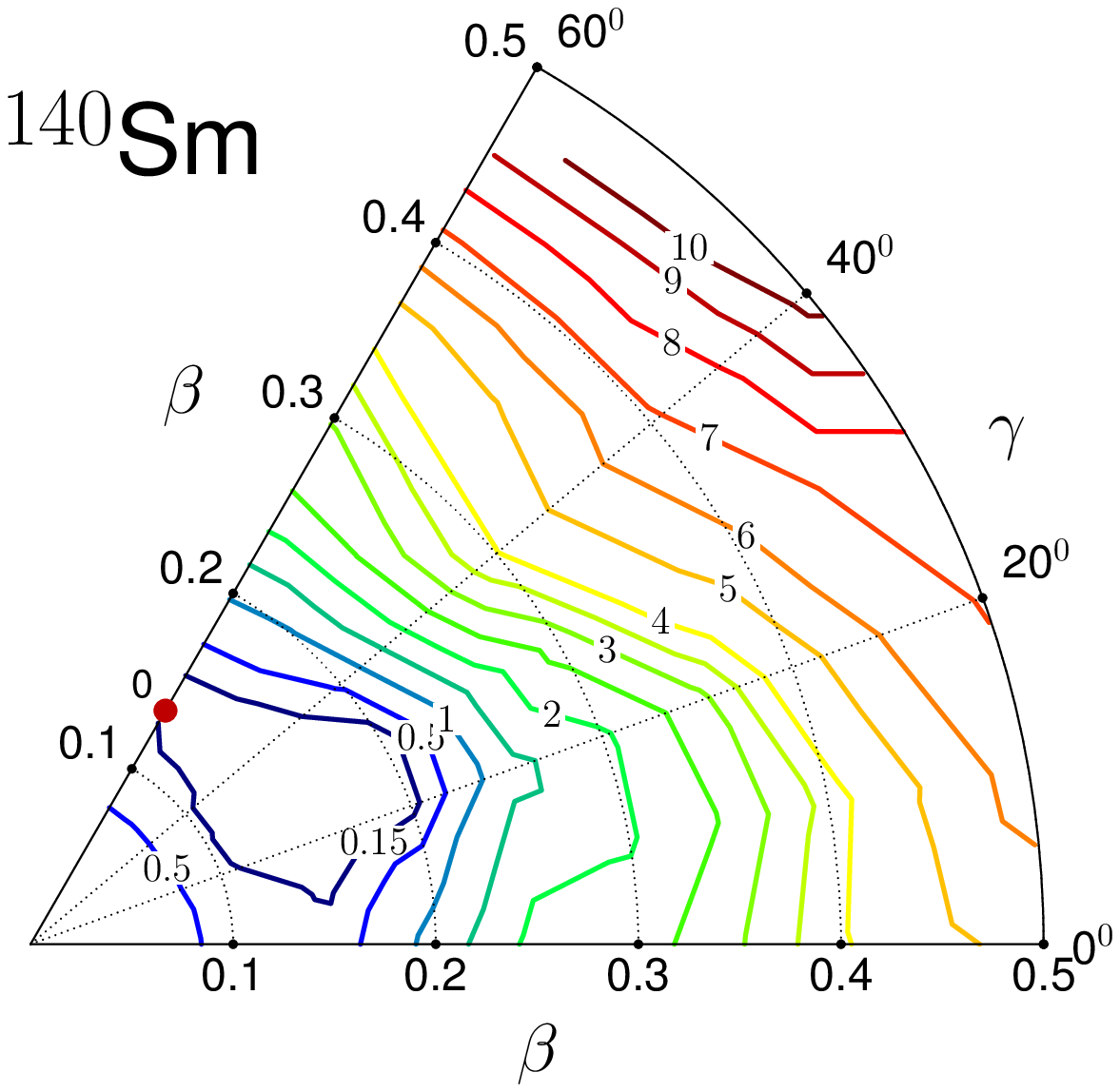}\\
\includegraphics[scale=0.45]{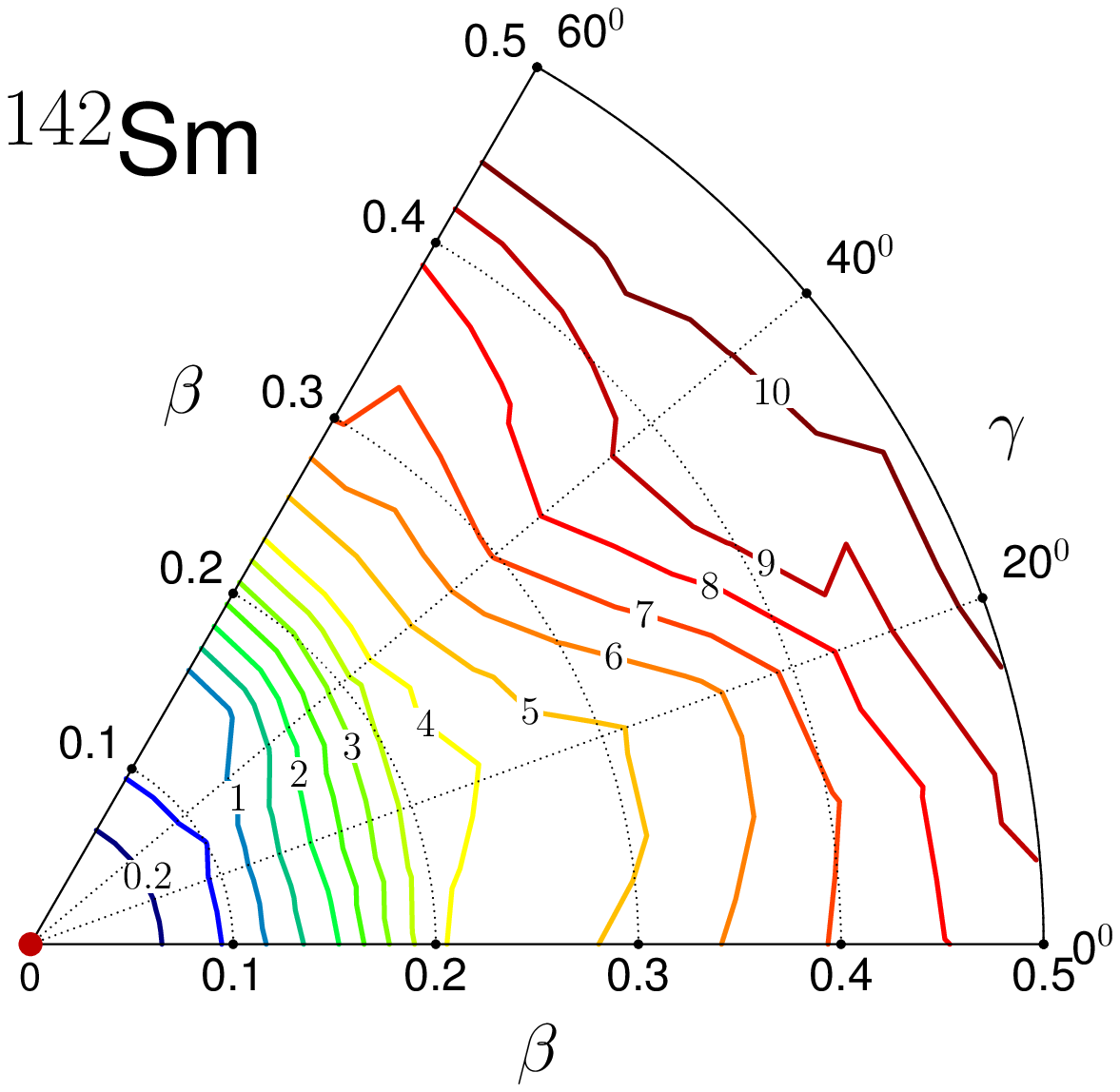} &
\includegraphics[scale=0.45]{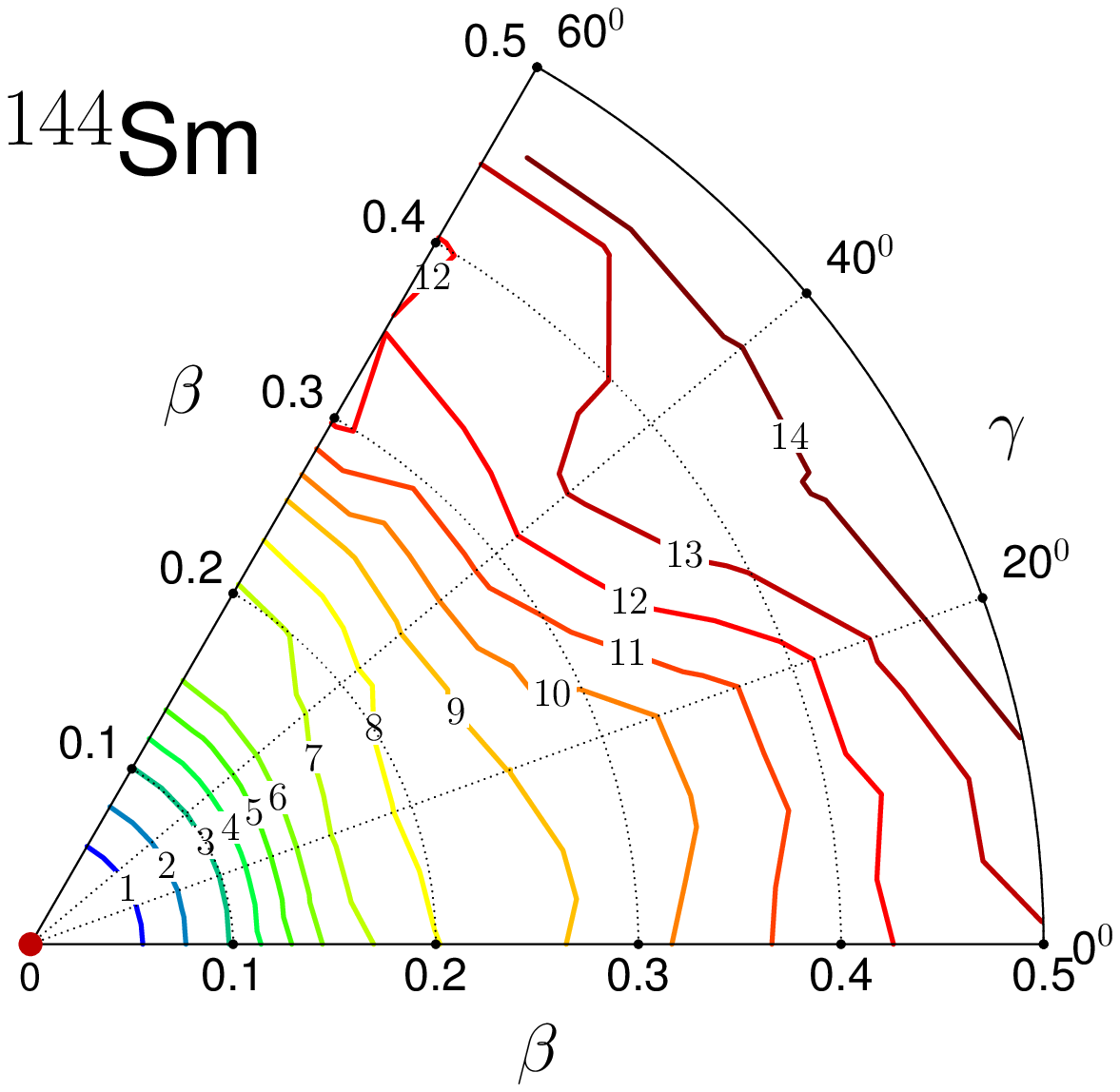}
\end{tabular}
\caption{\label{fig:pes_smA}(Color online)  Self-consistent RHB triaxial
quadrupole binding energy maps of the even-even isotopes $^{134-144}$Sm
in the $\beta-\gamma$ plane
($0\le \gamma \le 60^0$). All energies are normalized with respect to the binding
energy of the absolute minimum (red dot). The contours join points on
the surface with the same energy (in MeV).}
\end{figure}

\begin{figure}
\begin{tabular}{cc}
\includegraphics[scale=0.45]{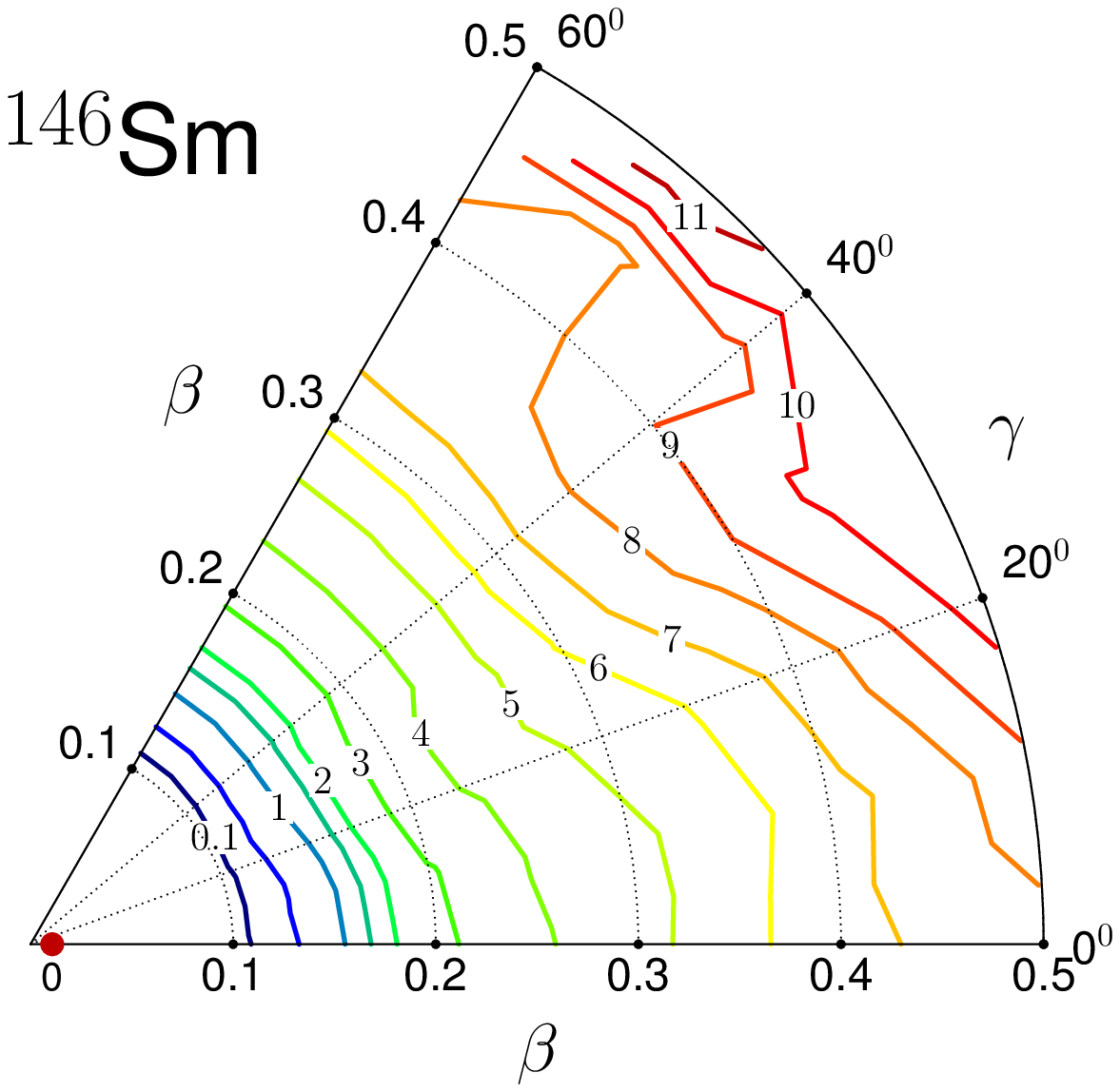}&
\includegraphics[scale=0.45]{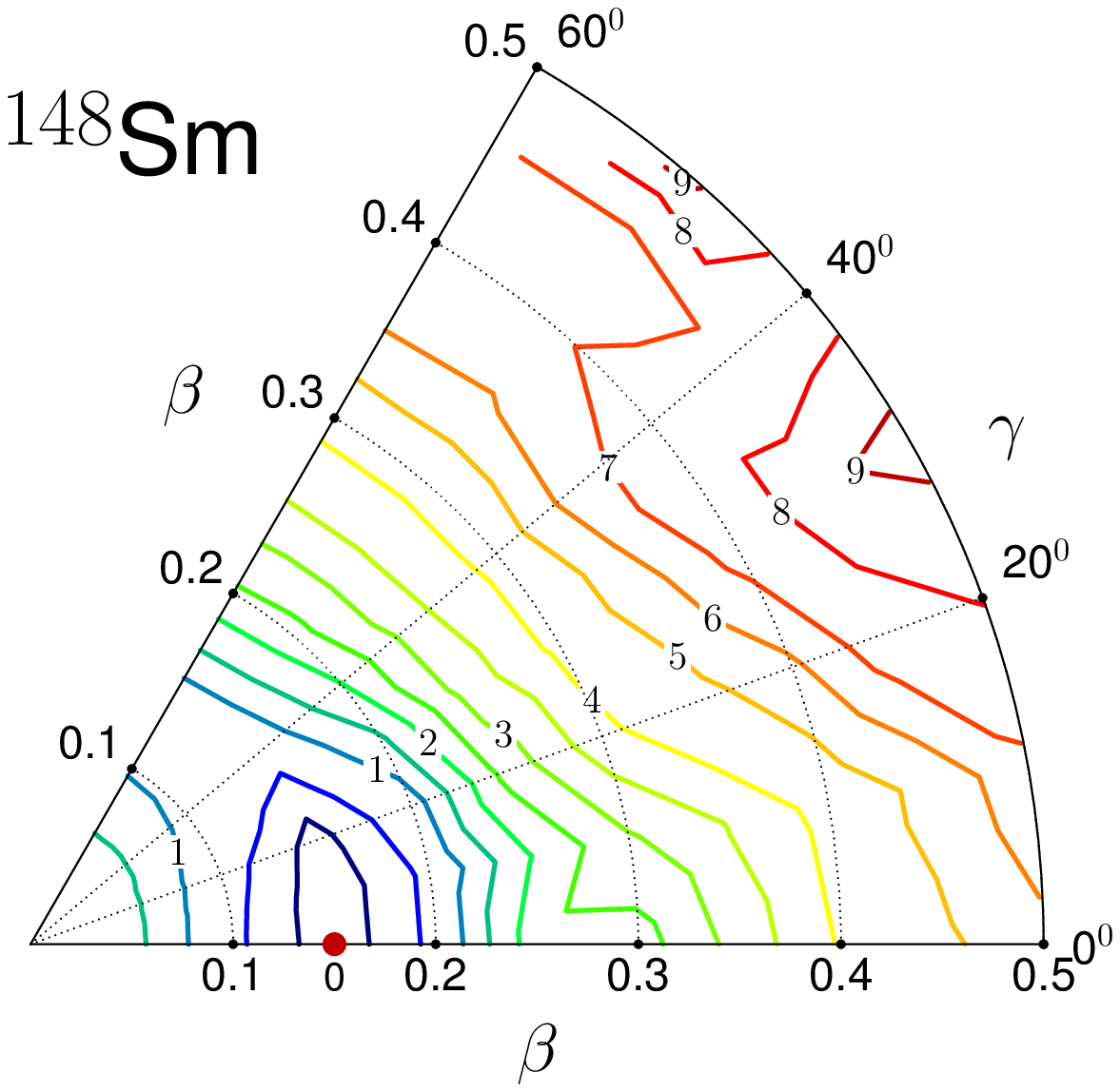}\\
\includegraphics[scale=0.45]{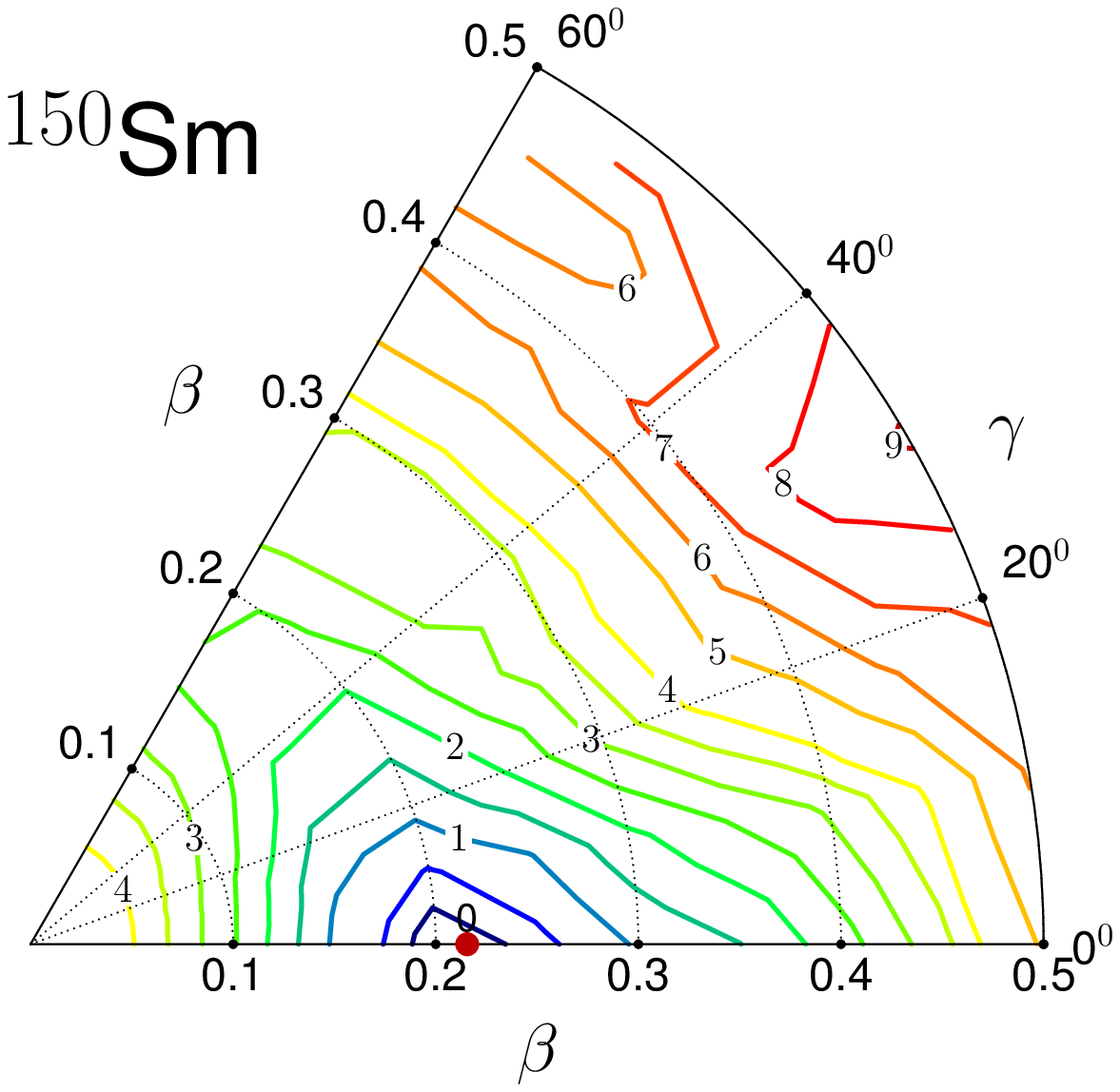} &
\includegraphics[scale=0.45]{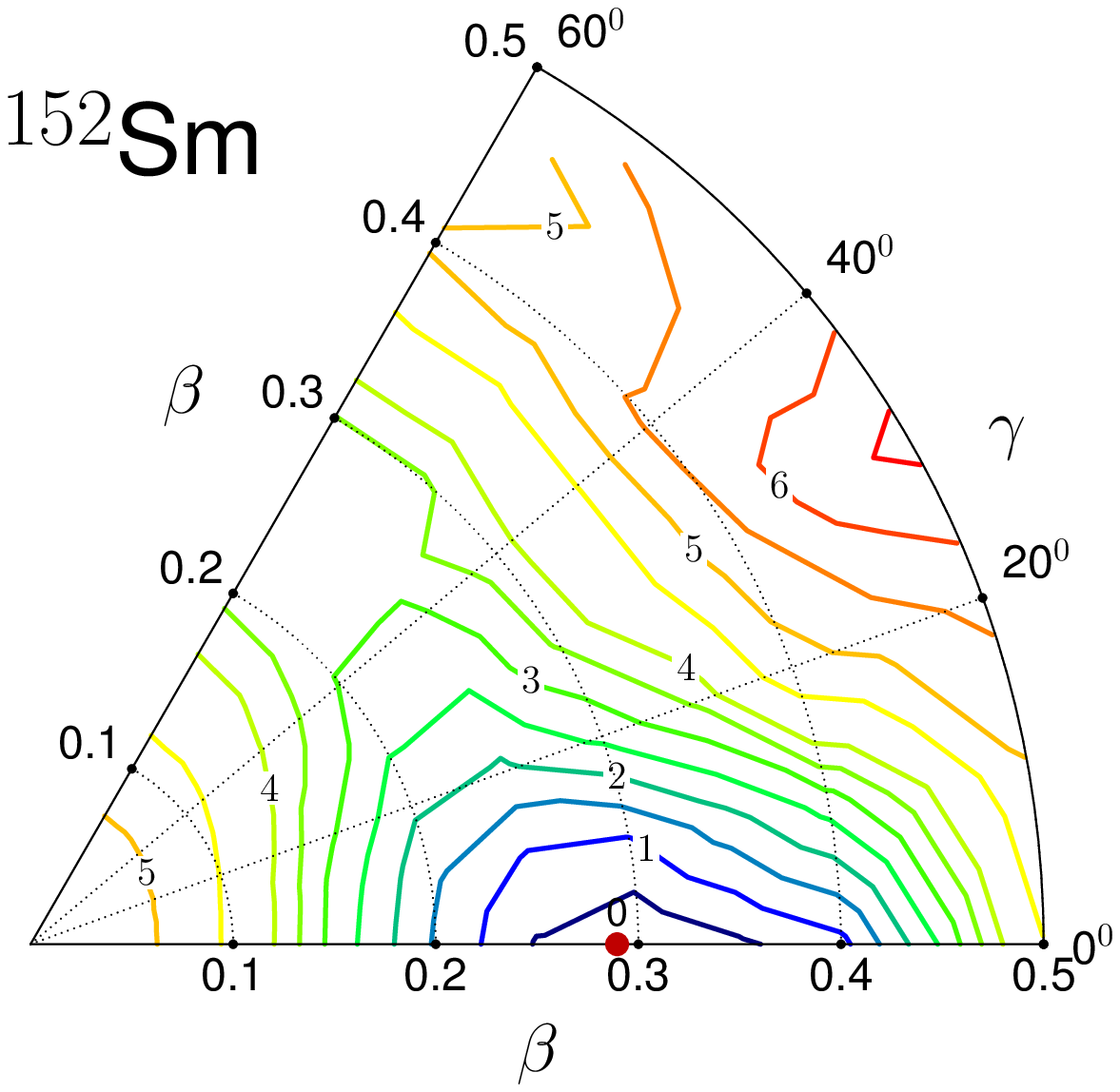}\\
\includegraphics[scale=0.45]{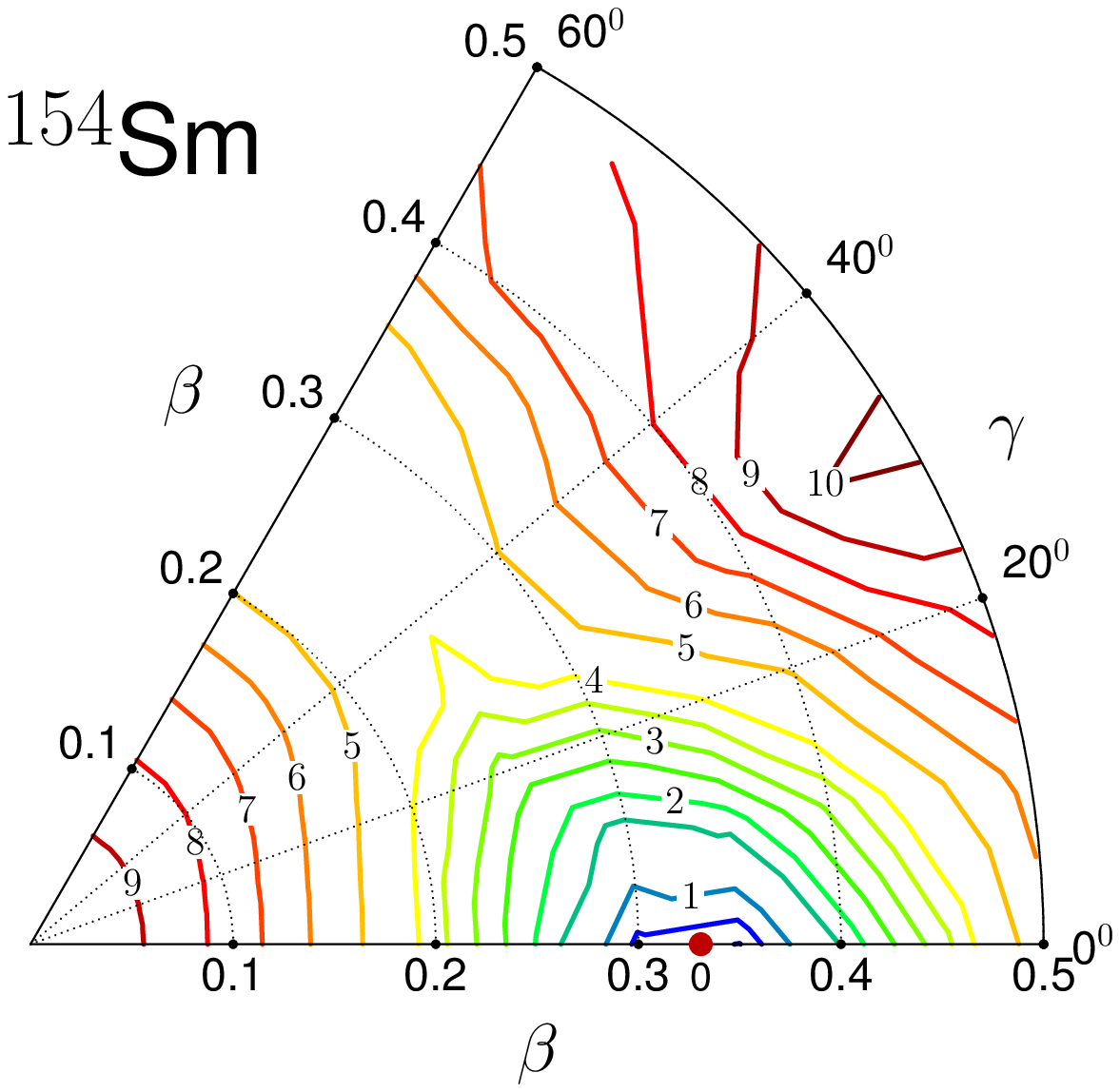} &
\includegraphics[scale=0.45]{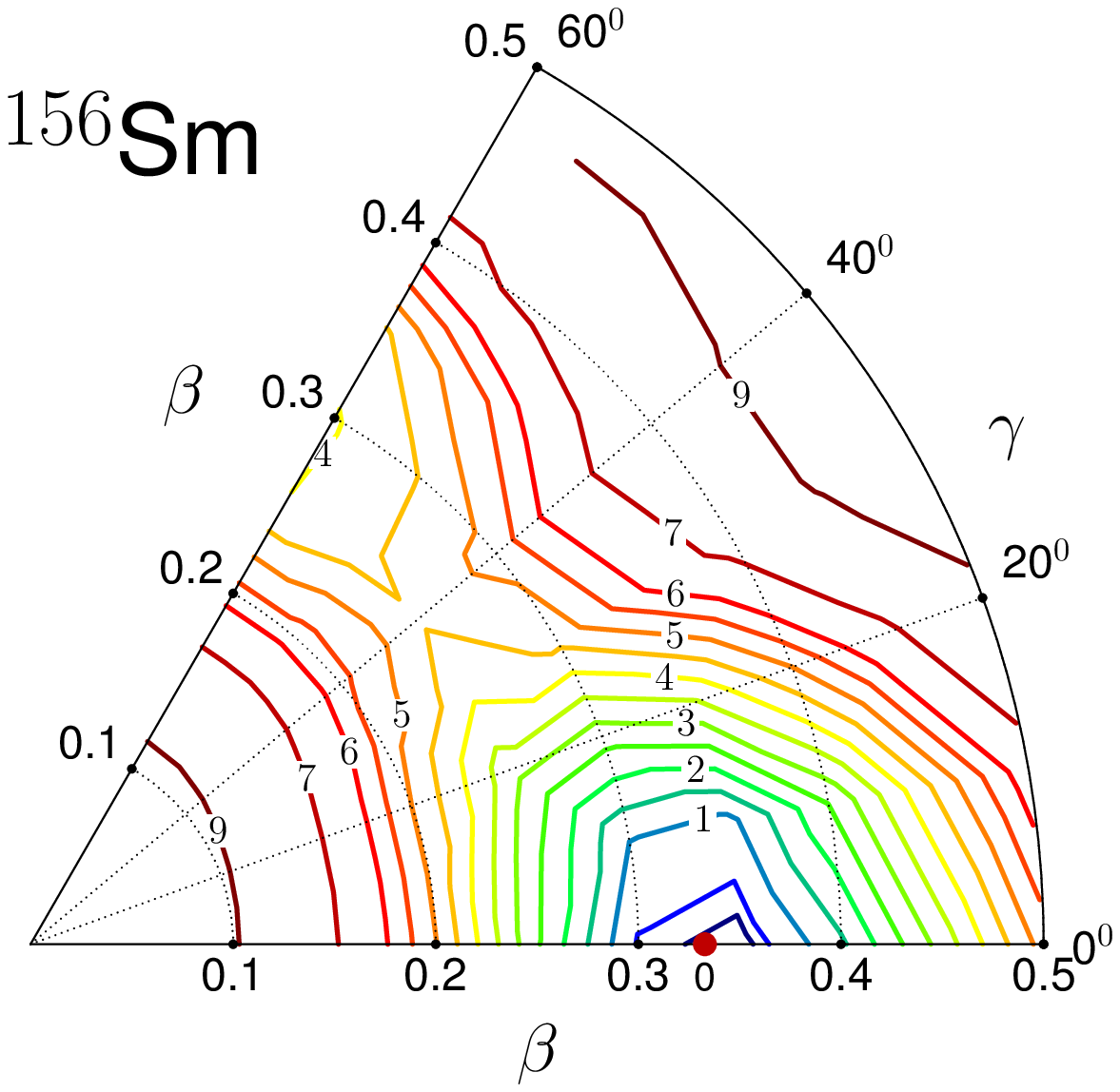}
\end{tabular}
\caption{\label{fig:pes_smB}(Color online) Same as Fig.~\ref{fig:pes_smA},
but for the isotopes $^{146-156}$Sm. }
\end{figure}

In Figs.~\ref{fig:pes_smA} and \ref{fig:pes_smB} we display the
self-consistent RHB triaxial quadrupole binding energy maps of the
$^{134-156}$Sm isotopes in the $\beta-\gamma$ plane ($0^0\le \gamma
\le 60^0$). The map of the energy surface as a function of the
quadrupole deformation is obtained by imposing constraints on the
axial and triaxial quadrupole moments. The method of quadratic
constraint uses an unrestricted variation of the function
\begin{equation}
\langle \hat{H}\rangle +\sum_{\mu=0,2}{C_{2\mu}
   \left(\langle \hat{Q}_{2\mu}\rangle -q_{2\mu} \right)^2},
\end{equation}
where $\langle \hat{H}\rangle$ is the total energy, and $\langle
\hat{Q}_{2\mu}\rangle$ denotes the expectation value of the mass
quadrupole operators:
\begin{equation}
\label{eq:quadrupole-constraints}
\hat{Q}_{20}=2z^2-x^2-y^2 \quad \textrm{and} \quad \hat{Q}_{22}=x^2-y^2 \;.
\end{equation}
$q_{2\mu}$ is the constrained value of the multipole moment, and
$C_{2\mu}$ the corresponding stiffness constant~\cite{RS.80}.

The energy maps shown in Figs.~\ref{fig:pes_smA} and
\ref{fig:pes_smB} nicely illustrate the gradual transition from the
prolate and  $\gamma$-soft deformed light isotopes $^{134,136}$Sm,
through the spherical $N=82$ neutron closed-shell nucleus $^{144}$Sm,
to the strongly prolate deformed, axial nuclei $^{154,156}$Sm. The
isotopes below the $N=82$ closed shell are all $\gamma$-soft and,
just before the shell closure, one finds a slightly oblate minimum in
$^{140}$Sm. Sm nuclei with $N > 82$ quickly develop a pronounced
prolate deformation, much stiffer with respect to the $\gamma$ degree
of freedom than isotopes below $N=82$. Heavy Sm isotopes are
characterized by axially symmetric shapes with pronounced prolate
minima at $\beta > 0.3$.

\begin{figure}
\begin{tabular}{cc}
\includegraphics[scale=0.45]{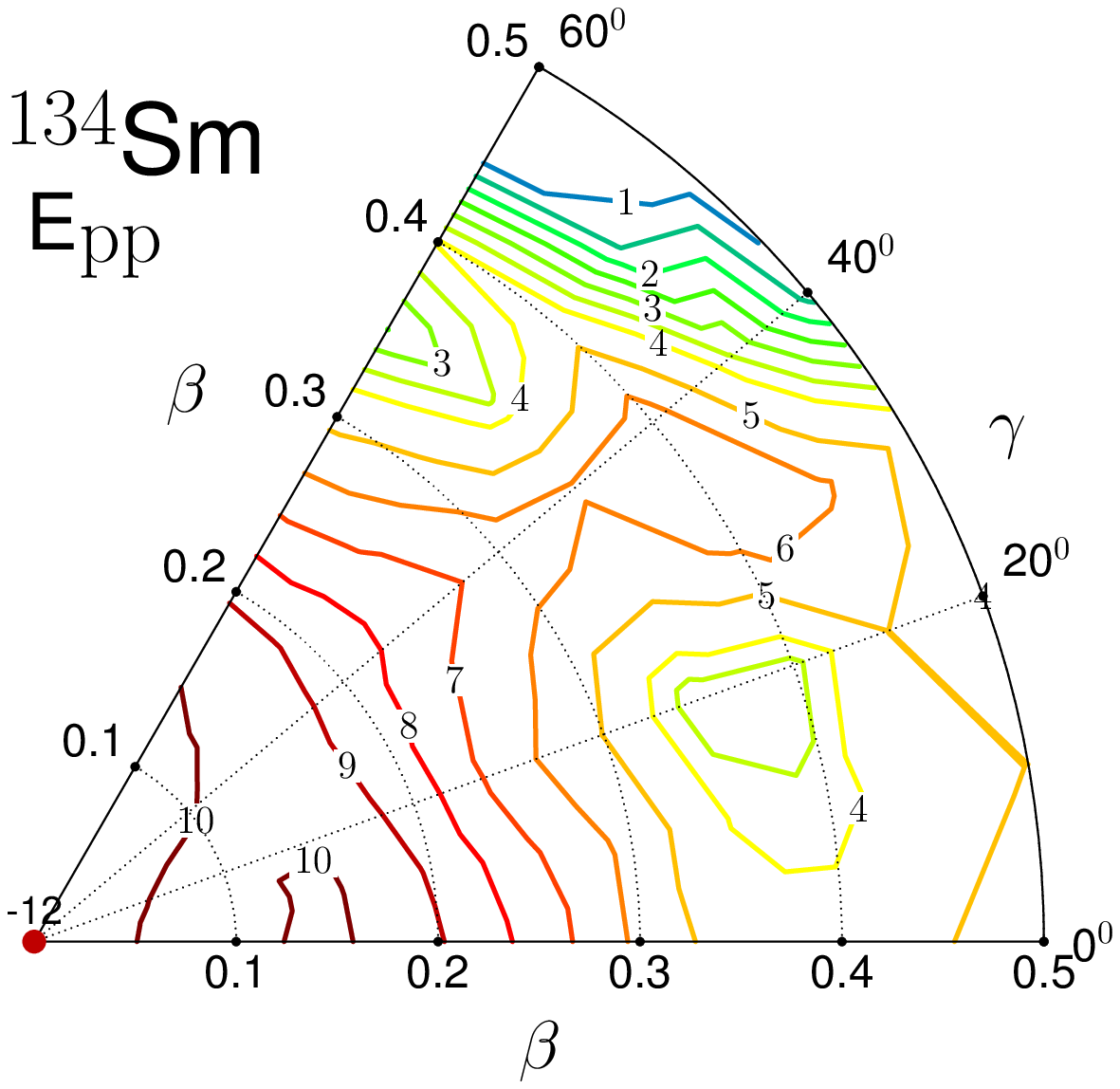}&
\includegraphics[scale=0.45]{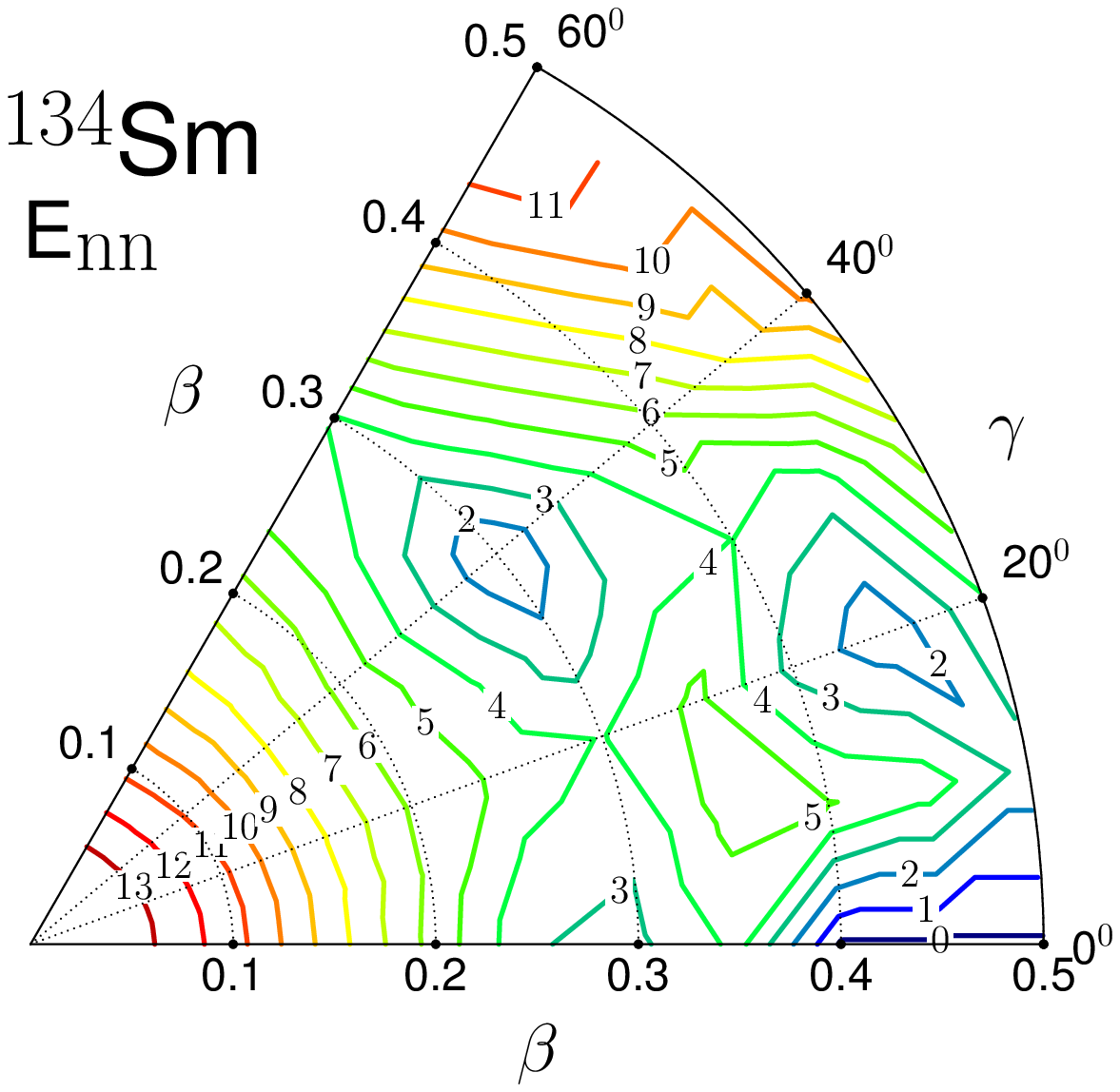}\\
\includegraphics[scale=0.45]{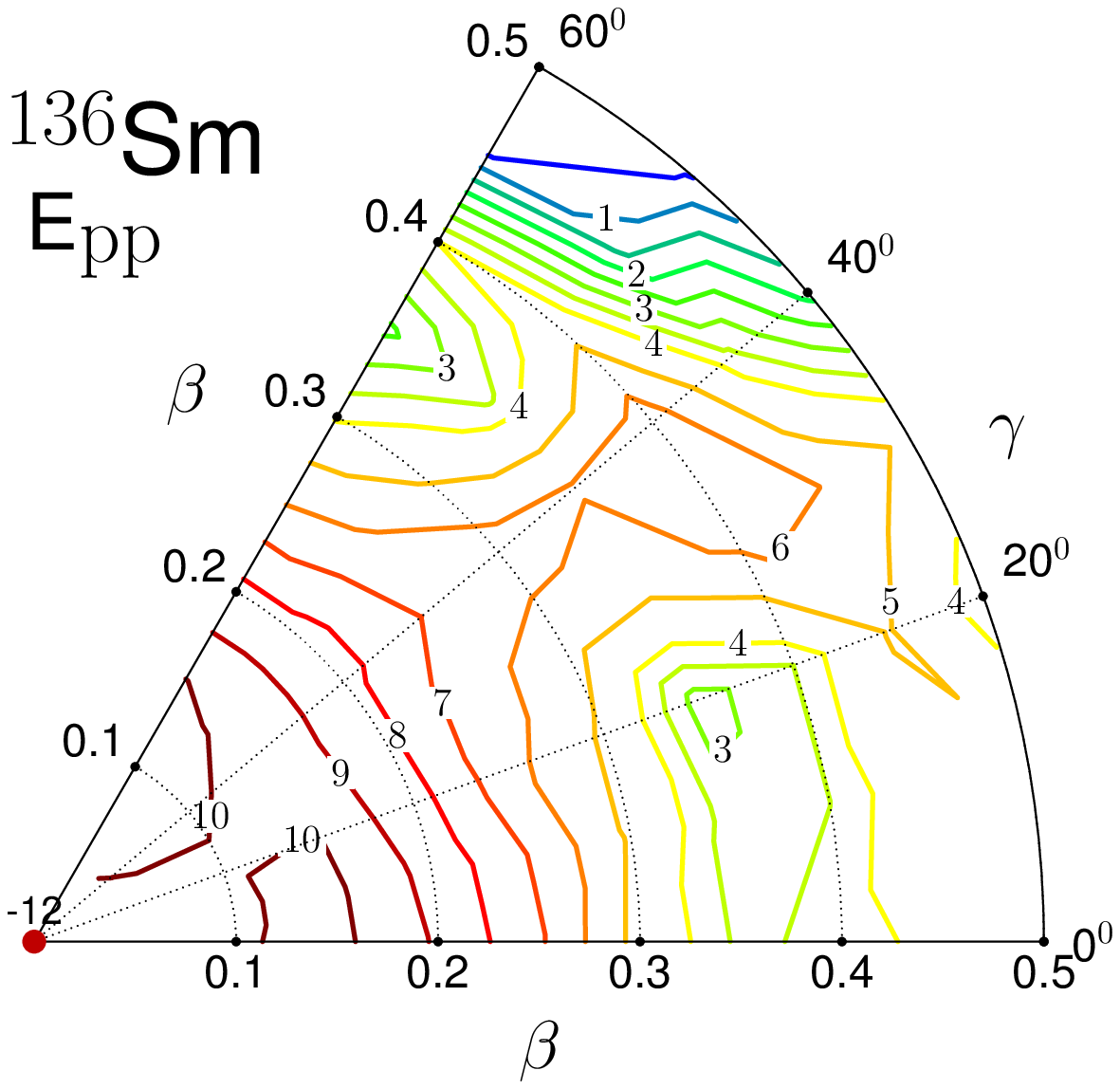} &
\includegraphics[scale=0.45]{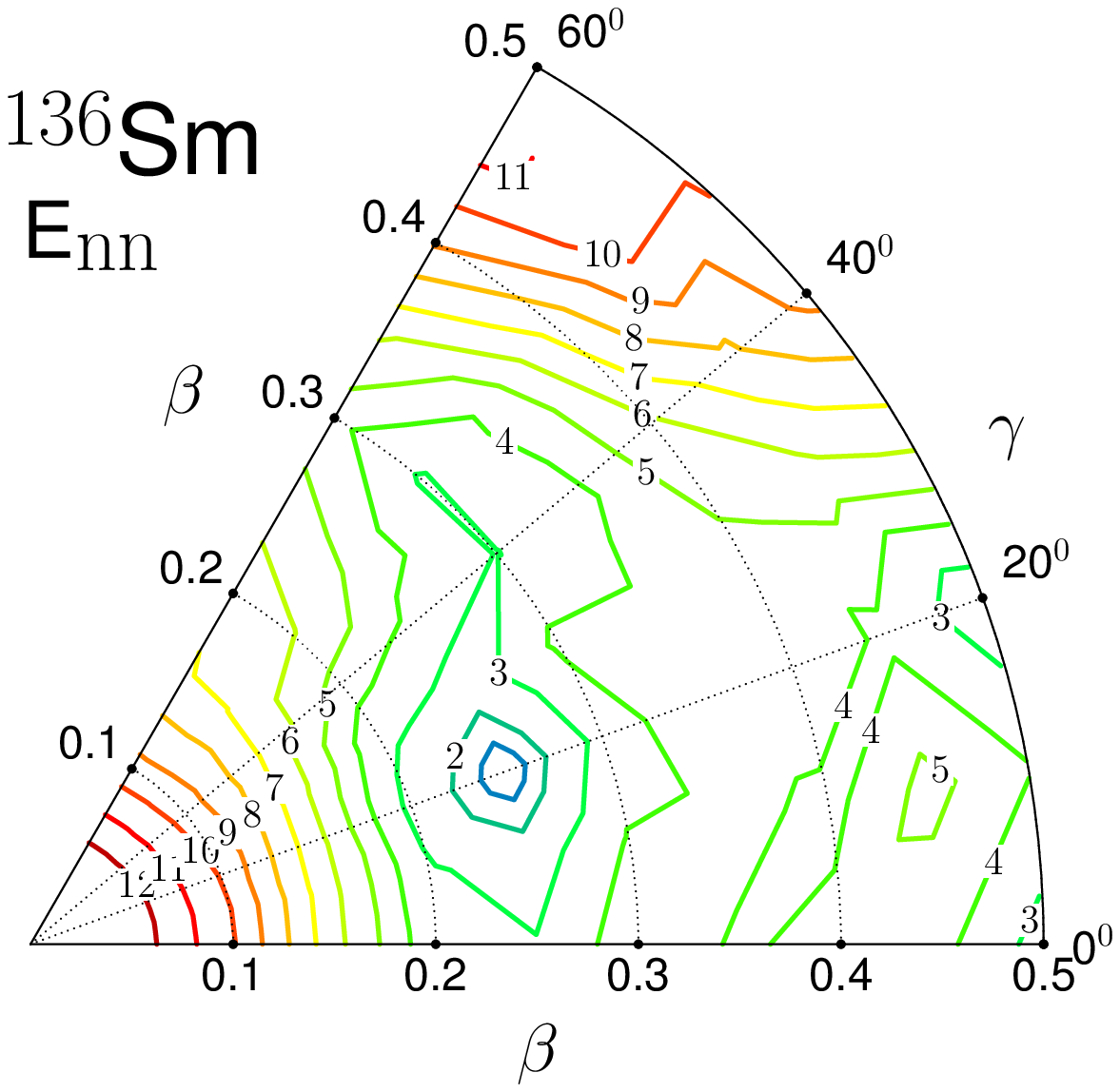}\\
\includegraphics[scale=0.45]{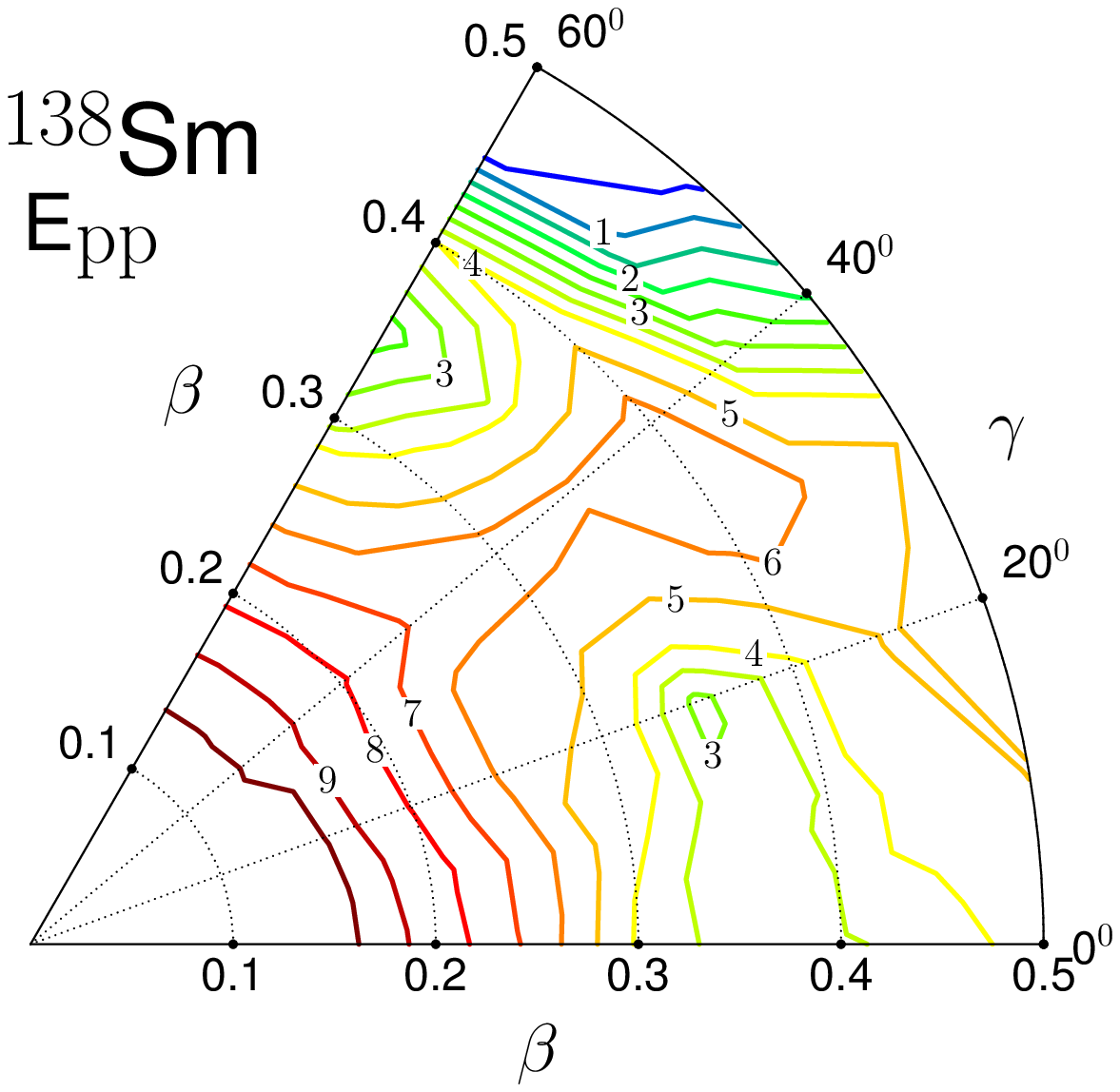} &
\includegraphics[scale=0.45]{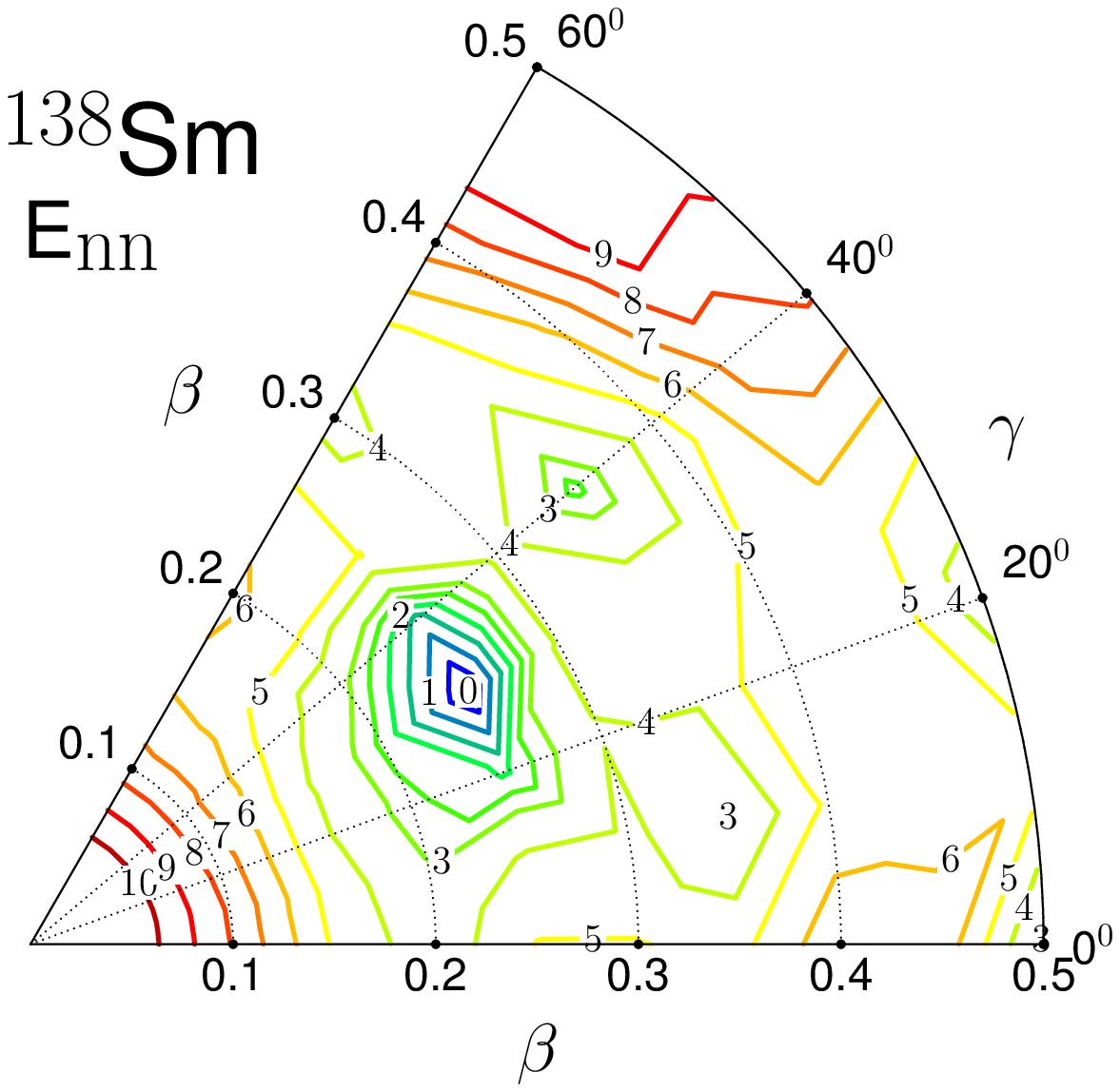}
\end{tabular}
\caption{\label{fig:pairing_smA}(Color online)
Proton (left column) and neutron (right column) pairing energies
of $^{134,136,138}$Sm in the
$\beta - \gamma$ plane ($0\le \gamma\le 60^0$).
The contours join points
on the surface with the same energy (in MeV).}
\end{figure}

\begin{figure}
\begin{tabular}{cc}
\includegraphics[scale=0.45]{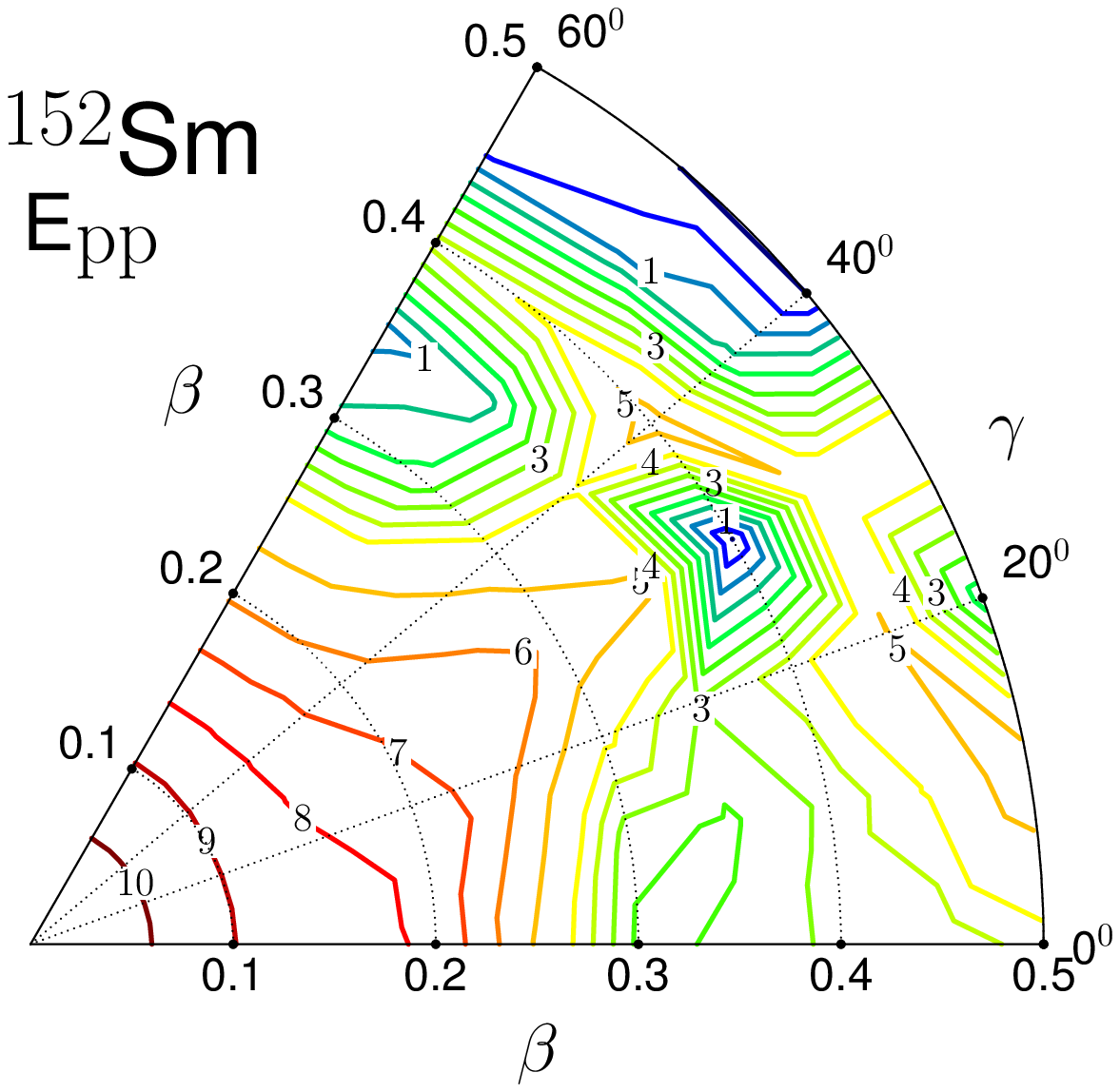}&
\includegraphics[scale=0.45]{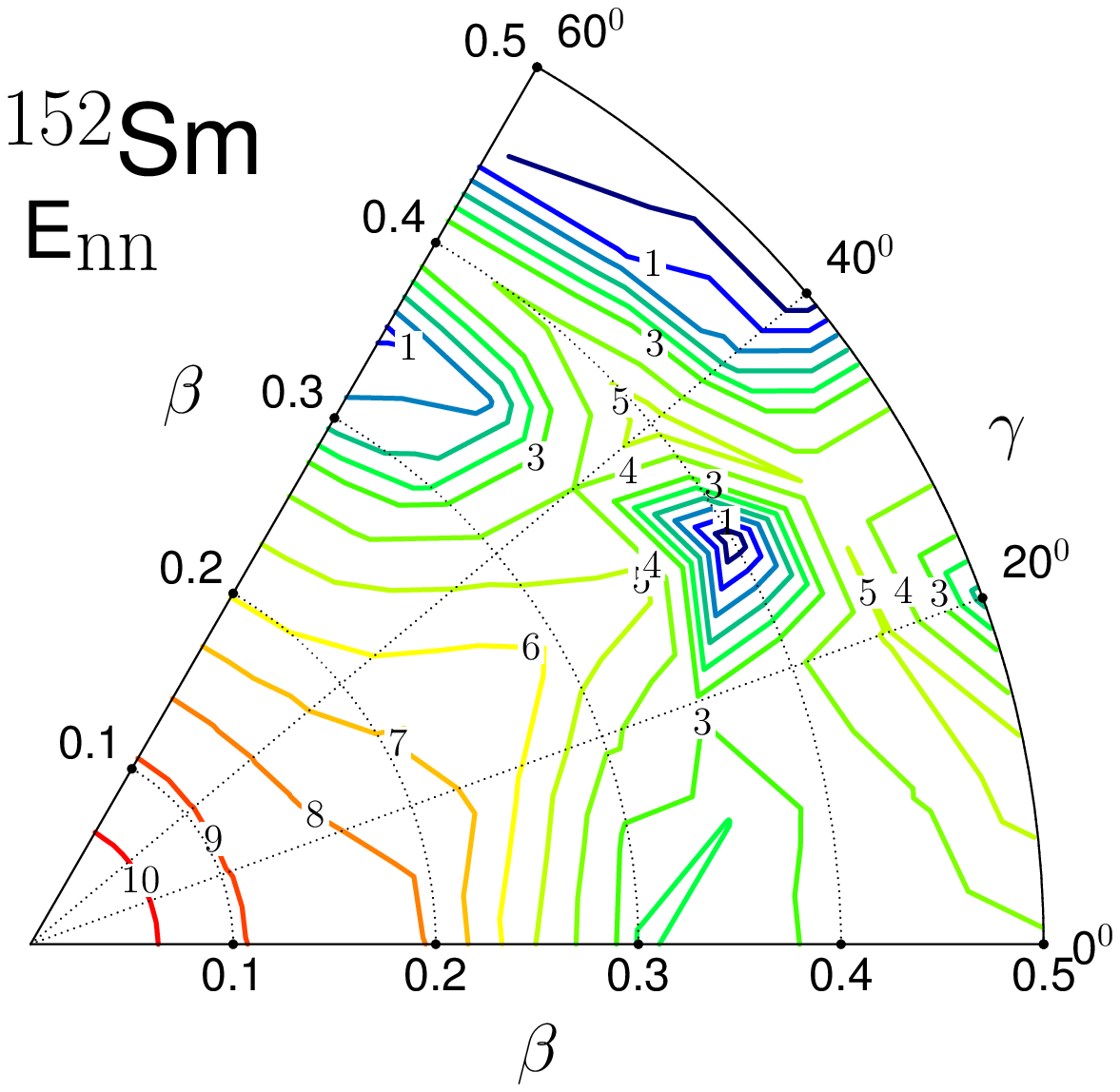}\\
\includegraphics[scale=0.45]{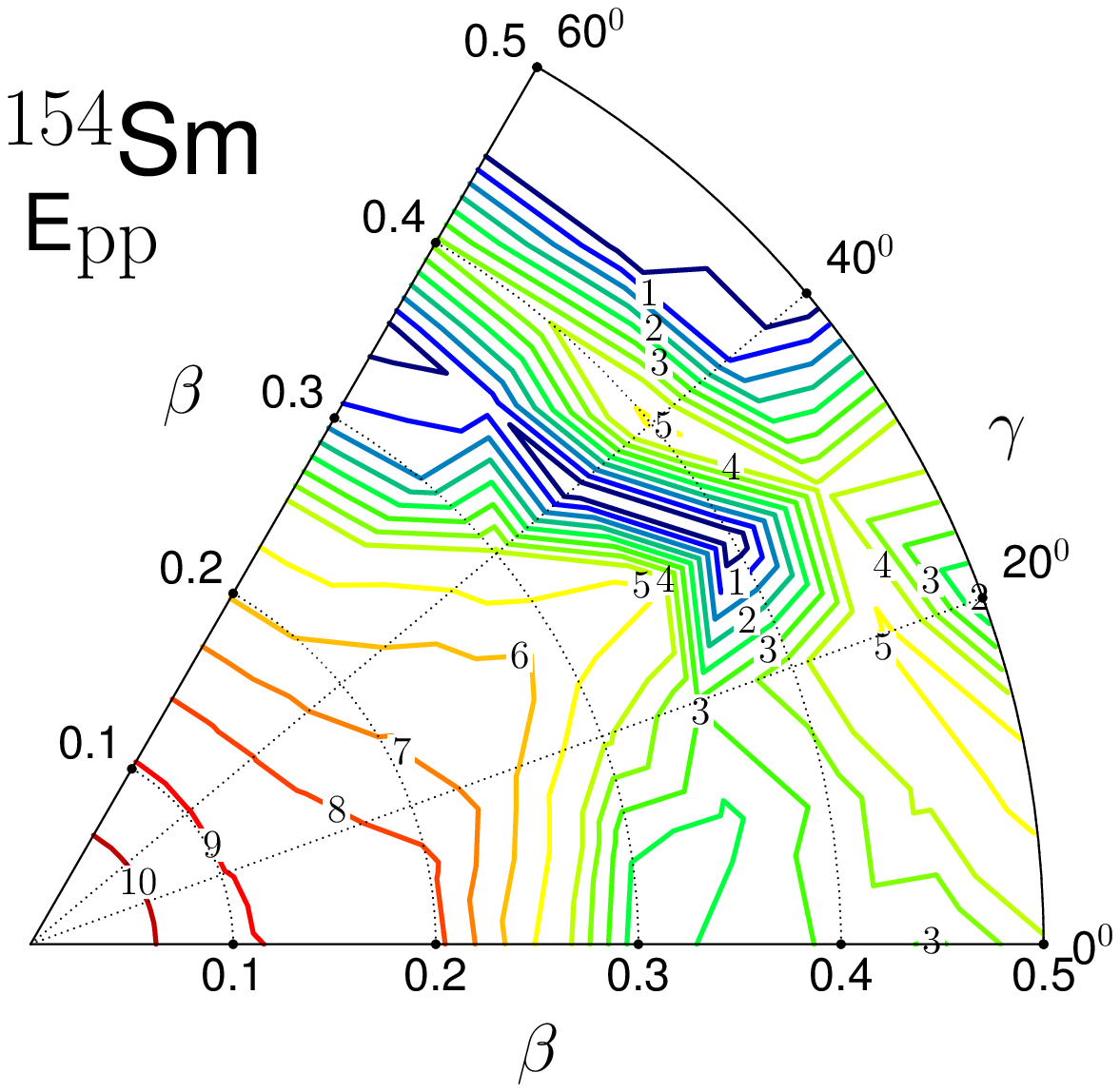} &
\includegraphics[scale=0.45]{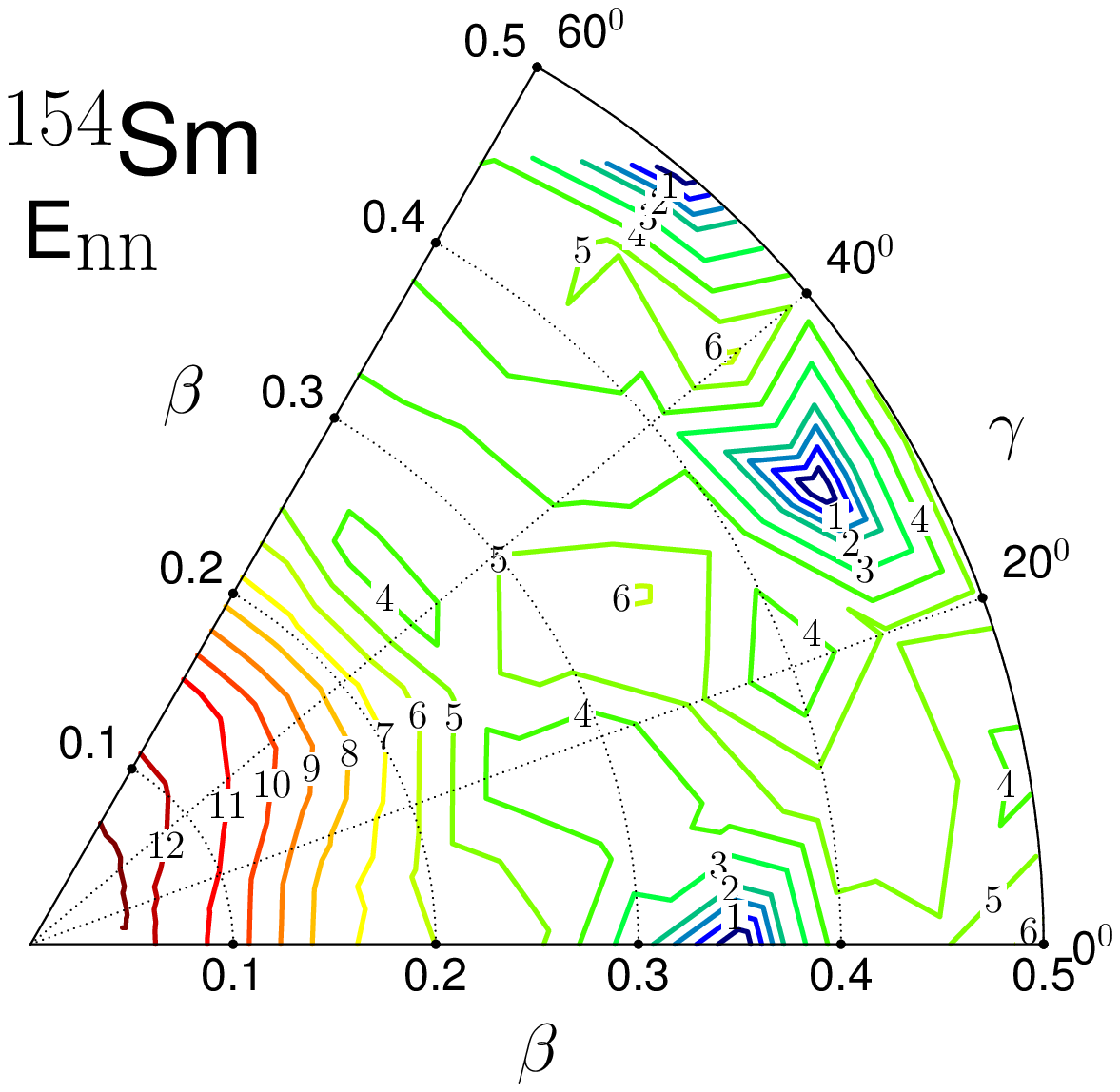}\\
\includegraphics[scale=0.45]{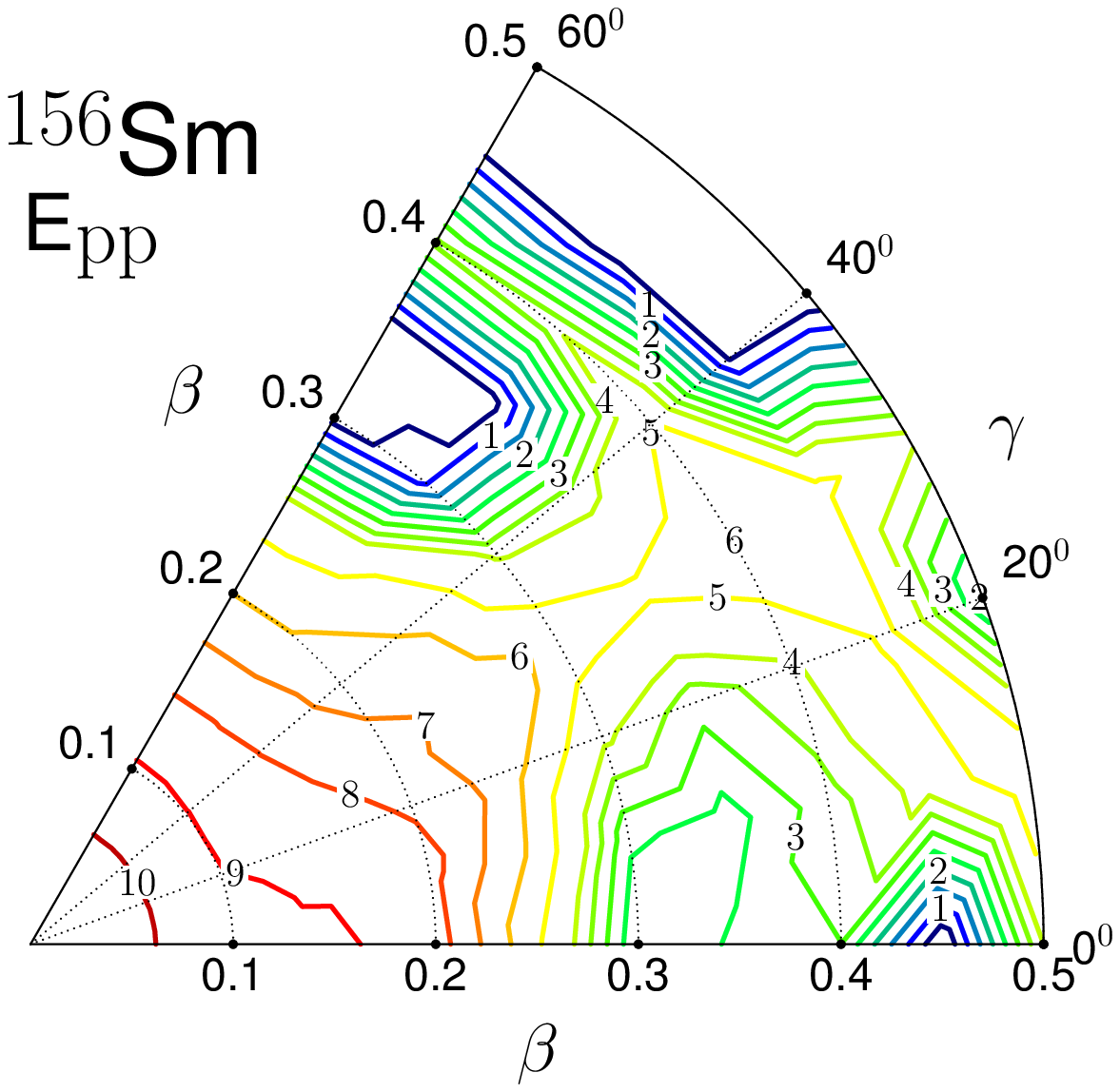} &
\includegraphics[scale=0.45]{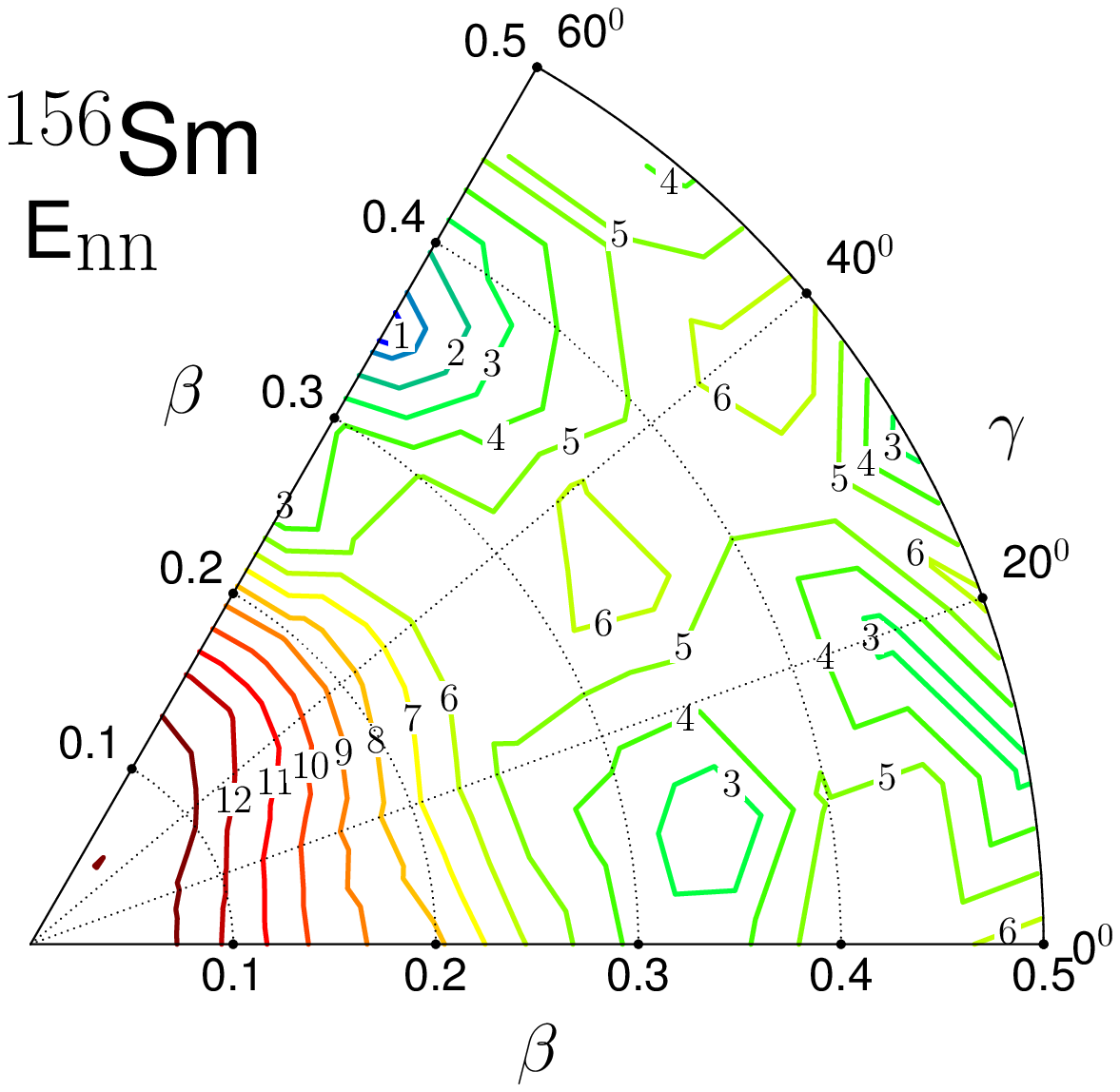}
\end{tabular}
\caption{\label{fig:pairing_smB}(Color online) Same as Fig.~\ref{fig:pairing_smA},
but for the isotopes $^{152,154,156}$Sm. }
\end{figure}

The binding energy maps correspond to self-consistent solutions of
the RHB equations, obtained by expanding the nucleon spinors and the
meson fields in the basis of a three-dimensional harmonic oscillator
(HO) in Cartesian coordinates. In the present calculation the basis
includes $N_f^{max} = 14$ major oscillator shells. In
Figs.~\ref{fig:pairing_smA} and \ref{fig:pairing_smB} we plot the
corresponding contour maps of the proton and neutron pairing energies
in the $\beta - \gamma$ plane for the three lighter isotopes
$^{134,136,138}$Sm, and for the three heavier nuclei
$^{152,154,156}$Sm, respectively. Using the separable pairing force
Eq.~(\ref{pp-force}), the pairing field Eq.~(\ref{eq:pairing-field})
is calculated as a sum of a finite number of separable terms. 3D
calculations in the $pp$ channel have been checked by comparing the
results for the ground state properties of $^{134-154}$Sm, with those
obtained using the axial RHBZ code with the same
separable pairing force~\cite{TMR.09c}, and with the pairing 
part of the original Gogny force~\cite{BGG.91,LVRS.99}. 
In the case when axial symmetry is assumed, the
expansion for the pairing field runs over the quantum numbers $N_z$
and $N_p$ of the HO in the center-of-mass frame, corresponding to the
$z$ and $\rho$ coordinates of the cylindrical coordinate system:
\begin{equation}
\label{eq:expansion-RHBZ}
\Delta_{12} = -G\sum_{N_z}^{N_z^0}\sum_{N_p}^{N_p^0}W_{12}^{N_zN_p}
   P_{N_z}P_{N_p}.
\end{equation}
The maximal values for the quantum numbers in the expansion of Dirac
spinors are $n_z^0=N_f^{max}$ and $n_p^0=N_f^{max}/2$, i.e. the
maximal values for the coefficients in  expansion
(\ref{eq:expansion-RHBZ}) are $N_z^0=2n_z^0=2N_f^{max}$ and
$N_p^0=2n_p^0=N_f^{max}$. In Ref.~\cite{TMR.09c} it has been shown
that, for axial calculations of prolate deformed nuclei, sufficient
accuracy is achieved if the expansion of pairing matrix elements is
limited to: $ N_p \le N_p^0=5$, and $N_z\le N_z^0=14$. For this
choice of the cut-off in the expansion of the pairing matrix elements
in the basis of the HO in the center-of-mass frame, the resulting
pairing energies reproduce to a very good approximation results
obtained with the calculation in the full basis, and also those
obtained with the Gogny force D1S in the pairing channel. In the
present 3D calculation we have, therefore, imposed the following
cut-off condition for the expansion in
Eqs.~(\ref{eq:matrix-element}), (\ref{eq:pairing-field}) and
(\ref{eq:separable-parameters}):
\begin{equation}
N_z =0,2,\dots,N_z^0=14\quad \textnormal{and} \quad
N_x+N_y =0,2,\dots,2N_p^0=5.
\end{equation}

\begin{figure}
\centering
\includegraphics[scale=0.55]{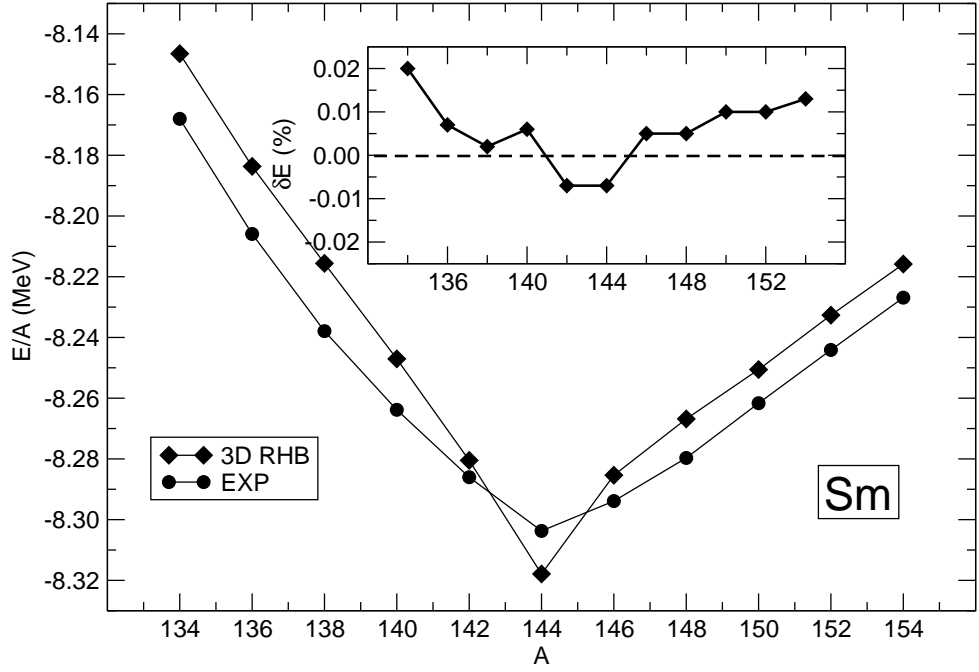}
\caption{\label{figA} Binding energy per nucleon for the sequence of
Sm isotopes, calculated with the 3D RHB model and compared to
data~\cite{AWT.03}. In the inset we display the relative differences
(in percent):
$(E^{\textrm{RHBZ}}-E^{\textrm{3DRHB}})/E^{\textrm{3DRHB}}$, between
the binding energies calculated using the 3D RHB and the axial (RHBZ)
relativistic Hartree-Bogoliubov models. }
\end{figure}

\begin{figure}
\centering
\includegraphics[scale=0.55]{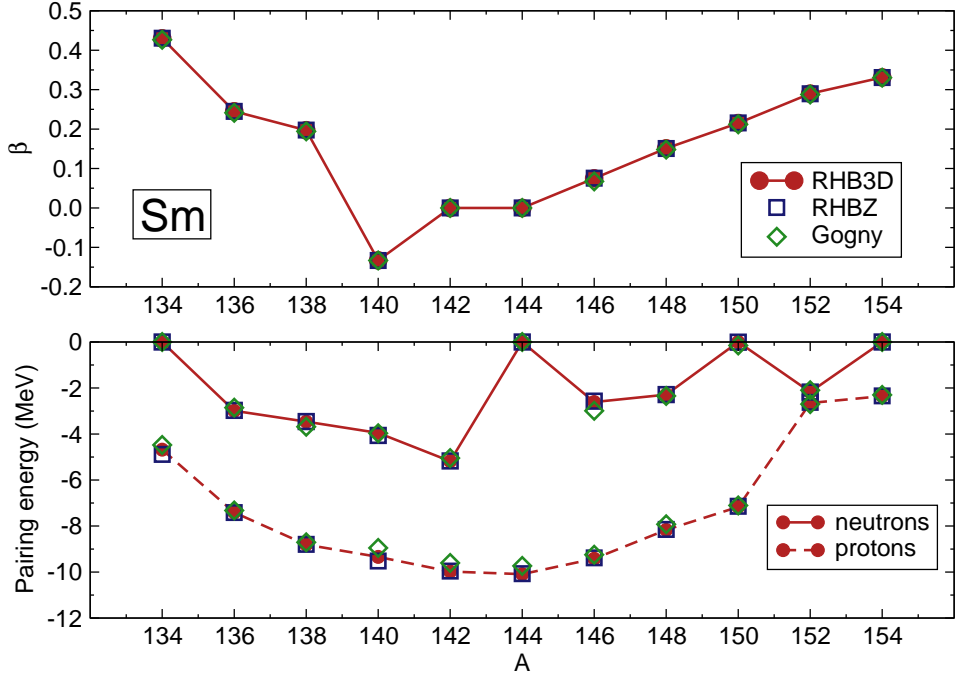}
\caption{\label{figC} (Color online) 3D RHB (filled circles) and axial (RHBZ) 
empty symbols) results for the self-consistent ground-state quadrupole
deformations (upper panel), and neutron and proton pairing energies
(lower panel) of even-A Sm isotopes. In calculations with axial symmetry both the
separable force (squares) and the Gogny D1S force~\cite{BGG.91}
(diamonds) are used.}
\end{figure}

In Fig.~\ref{figA} we display the 3D RHB ground-state binding
energies for the Sm isotopes ($72\le N\le 92$), in comparison with
data from the compilation of Audi and Wapstra~\cite{AWT.03}.
Calculations have also been performed with the axial RHBZ
code~\cite{TMR.09c}, and the inset we plot the relative differences
(in percent):
$(E^{\textrm{RHBZ}}-E^{\textrm{3DRHB}})/E^{\textrm{3DRHB}}$, between
the corresponding ground-state binding energies. As a further test,
Fig.~\ref{figC} compares the 3D RHB and axial (RHBZ) 
results for the self-consistent ground-state
quadrupole deformations, neutron and proton pairing energies of
even-A Sm isotopes.  In calculations with axial symmetry (RHBZ) both the
separable force and the Gogny D1S force~\cite{BGG.91}  are used in the pairing channel.
The excellent agreement between the three sets of results demonstrate not only the 
numerical accuracy of the new 3D computer code, but also show that
using the separable pairing force in deformed nuclei, virtually identical pairing energies 
are calculated as with the original Gogny
force. 

\section{\label{conclusions}Summary and Outlook}

Realistic self-consistent mean-field calculations based on
finite-range interactions, including exchange terms and/or pairing
correlations, still present a considerable computational challenge 
\cite{DGG.06,SE.08,HG.07,LRG.09}, particularly if one considers 
complex triaxial shapes or extensions beyond the 
simplest mean-field approximation. A great advantage of 
mean-field models based on Skyrme -like zero-range interactions 
is that they provide a simple and elegant treatment of Fock exchange 
and pairing terms \cite{VB.72,DFT.84}. The disadvantage of such forces, 
i.e. the fact that they are constant in momentum space and can induce 
scattering of nucleons very high up into the continuum, does not 
appear at the Hartree or Hartree-Fock level, at which one 
considers only momenta up to the Fermi surface.  
In the pairing channel, however, because of the specific form
of the BCS or Bogoliubov ansatz that takes into account 
pairing correlations on the mean-field level, ultraviolet divergencies occur for
zero-range forces. One possible solution present the 
various cutoff procedures that have been used in the
literature (cf. Ref.~\cite{KAL.09} and references therein). These 
approximations all include additional non-physical cut-off parameters. 
This does not cause any problem in investigations along the valley of 
beta-stability, where gap parameters can be deduced from experimental masses. 
However, the use of cut-off parameters limits the predictive power of such models in 
unknown regions of the nuclear chart,  such as for superheavy elements or very 
neutron rich isotopes.

A completely different approach to the treatment of 
pairing correlations is the use of separable forces. A separable form 
of the pairing force for RHB calculations in finite nuclei 
has recently been introduced \cite{TMR.09a}. The force is separable in
momentum space, and is completely determined by two parameters that
are adjusted to reproduce in symmetric nuclear matter the bell-shape
curve of the pairing gap of the Gogny force. Because of translational invariance, 
the pairing force is no longer exactly separable in coordinate space, but
Talmi-Moshinsky techniques allow a simple transformation into a 
quickly converging series of separable terms in a harmonic oscillator basis.
Although different from the Gogny force, the corresponding
effective pairing interaction has been shown to reproduce with 
high accuracy pairing gaps and energies calculated with 
the original Gogny force,  both in spherical and axially deformed
nuclei. In particular, this approach retains the basic 
advantage of the finite-range Gogny force, i.e. the 
natural cut-off in momentum space.

Applications have so far been restricted to the description
of spherical~\cite{TMR.09a} and  axially deformed nuclei~\cite{TMR.09c}.
In this work we have extended the model to describe triaxially deformed nuclei.
The numerical accuracy of the new model has been analyzed by comparing 
results with those obtained in axially-symmetric calculations, using both the 
separable force as well as the original
Gogny D1S force in the pairing channel.

To illustrate the applicability of this force in the description of realistic triaxial systems, 
we have explored the chain of Sm isotopes with
$Z=62$ protons, ranging from $^{134}$Sm to $^{156}$Sm. For the magic
neutron number $N=82$, i.e. for the $^{144}$Sm isotope, a  stable
spherical minimum is found in the $\beta-\gamma$ plane. The two
neighboring nuclei $^{142}$Sm and $^{146}$Sm are still spherical, but
with much softer energy surfaces. In heavier isotopes
we find a rather rapid transition to prolate shapes
with a well pronounced minima, and increasing 
$\beta$-deformation up to $\beta\approx 0.3$. 
In these heavier nuclei we also find a 
soft saddle point on the oblate side, that eventually 
becomes  a shallow second minimum in the isotope $^{156}$Sm. 
Decreasing the neutron number below the 
closed shell at $N=82$, a $\gamma$-soft valley develops with increasing 
$\beta$-deformation. For  $^{140}$Sm isotope we find a shallow oblate minimum,
whereas for the lighter isotopes the minima are located on the prolate side, but 
the calculation predicts large 
fluctuations in the $\gamma$-direction.

One can envisage many possible applications of the separable pairing force. 
The force is simple enough to be applied in otherwise
time consuming calculations, e.g. description of triaxial effects, rotating nuclei,
fission process, spherical and deformed QRPA. It can also be used in various
beyond mean-field extensions, such as restoration of broken symmetries,
fluctuations of quadrupole moment and particle-vibration coupling. 
In the current version of the model, the pairing force has been adjusted
to the pairing gap of the phenomenological Gogny D1S force. 
In the next step one could adjust the effective force to a 
pairing gap in nuclear matter calculated in a fully microscopic 
approach starting from inter-nucleon interactions \cite{KNV.05,GIF.09,Heb.09}. 

\bigskip \bigskip
\leftline{\bf ACKNOWLEDGMENTS} \noindent This work was supported in
part by MZOS - project 1191005-1010, and by the DFG cluster of
excellence \textquotedblleft Origin and Structure of the
Universe\textquotedblright\ (www.universe-cluster.de). T. N.
acknowledges support from the Croatian National Foundation for
Science. Y.T. and Z.Y.M. acknowledge support from the 
National Natural Science Foundation of China under Grants 10875150, 10775183, 
and 10535010, and the Major State Basis Research
Development of China under Contract 2007CB815000.
\bigskip

\appendix
\section{\label{App-basis}The single-nucleon basis }
The Dirac single-nucleon spinors are expanded in the basis of
eigenfunctions of a three-dimensional harmonic oscillator (HO) in
Cartesian coordinates. In one dimension:
\begin{equation}
\phi_{n_\mu}(x_\mu) = b_\mu^{-1/2} \mathcal{N}_{n_\mu} H_{n_\mu}(\xi_\mu)e^{-\xi_\mu^2/2}\;,
\quad (\mu\equiv x,y,z)
\end{equation}
$\xi_\mu \equiv x_\mu/b_\mu$, and the oscillator length is defined as
\begin{equation}
b_\mu = \sqrt{\frac{\hbar}{m\omega_\mu}}\; .
\end{equation}
The normalization factor reads
\begin{equation}
\mathcal{N}_{n} = \pi^{-1/4}\left( 2^n n! \right)^{-1/2},
\end{equation}
and $H_n(\xi)$ denotes the Hermite polynomials~\cite{Abramowitz}
\begin{equation}
\label{orthogonality}
\int_{-\infty}^{\infty}{H_n(\xi)H_{n^\prime}(\xi)e^{-\xi^2}d\xi} = \delta_{nn^\prime}\;.
\end{equation}
The basis state can be defined as the product of three HO wave
functions (one for each dimension) and the spin factor:
\begin{equation}
\phi_{\alpha}(\mathbf{r};m_s) =
    \phi_{n_x}(\xi_x) \phi_{n_y}(\xi_y)\phi_{n_z}(\xi_z)\chi_{m_s},
\label{product}
\end{equation}
where the notation is: $\alpha \equiv \{n_x,n_y,n_z\}$. For each
combination of quantum numbers $\{n_x,n_y,n_z\}$, the spin part is
chosen in such a way that the basis state is an eigenfunction of the
x-simplex operator $\displaystyle \hat{S}_x=\hat{P}
e^{-i\pi\hat{J}_x}$, where $\hat{P}$ denotes the parity operator. The
positive and negative x-simplex operator eigenstates:
\begin{align}
\label{basis-pos}
\ket{n_x n_y n_z; i=+} &= \ket{n_xn_yn_z}\frac{i^{n_y}}{\sqrt{2}}
\left[\ket{\uparrow} -(-1)^{n_x}\ket{\downarrow}  \right], \\
\label{basis-neg}
\ket{n_x n_y n_z; i=-} &= \ket{n_xn_yn_z}(-1)^{n_x+n_y+1}\frac{i^{n_y}}{\sqrt{2}}
\left[\ket{\uparrow} +(-1)^{n_x}\ket{\downarrow}  \right],
\end{align}
are related by the time-reversal operator ($\hat{T}=i\sigma_y \hat{K}_0$)
\begin{equation}
\ket{n_x n_y n_z; i=-} = \hat{T} \ket{n_x n_y n_z; i=+} \; .
\end{equation}
For the Dirac spinor with positive simplex eigenvalue, the large
component corresponds to positive, and the small component to
negative eigenvalues
\begin{equation}
\psi_i(\mathbf{r},+) = \left( \begin{array}{c} f_i(\mathbf{r},+) \\
            ig_i(\mathbf{r},-) \end{array}  \right).
\end{equation}
The large and small component are expanded in terms of the basis
states Eqs.~(\ref{basis-pos}) and (\ref{basis-neg}):
\begin{equation}
f_i(\mathbf{r};+) =
    \sum_{\alpha}^{\alpha_{max}}{f_i^\alpha\phi_{\alpha}(\mathbf{r};+)}
\quad \textnormal{and} \quad
g_i(\mathbf{r};-) =
    \sum_{\bar{\alpha}}^{\bar{\alpha}_{max}}
             {g_i^{\bar{\alpha}}\phi_{\bar{\alpha}}(\mathbf{r};-)}\;.
\end{equation}
Positive simplex eigenstates are denoted by $\ket{\alpha}$, and
negative simplex eigenstates by $\ket{\bar{\alpha}}$. If the basis
states are arranged as:
$\{\alpha_1,\dots,\alpha_M,\bar{\alpha}_1,\dots \bar{\alpha}_M\}$,
the x-simplex operator has a simple block-diagonal form, whereas the
time-reversal operator is skew diagonal:
\begin{equation}
\hat{S}_x = i \left( \begin{array}{cc} \openone & 0 \\ 0 & -\openone
                    \end{array} \right) \quad \textnormal{and} \quad
\hat{T} =  \left( \begin{array}{cc} 0 & \openone \\  -\openone  & 0
                    \end{array} \right) \; .
\end{equation}
If the dimension of each simplex block ($i=\pm$) is $M$, the
dimension of the entire configuration space equals $2M$. In the
present implementation of the model parity is also conserved, and
this allows a further reduction of the basis to four simplex-parity
blocks. For a given maximal number of oscillator shells $N_{max}$,
the dimension of the HO basis can be determined as follows. The
states $\ket{n_x n_y n_z}$ within a major oscillator shell $N$ are
arranged as

\begin{center}
\begin{tabular}{ccc}
$n_x$     & $n_y$       &  $n_z$ \\ \hline \hline
  0           &    0           &  $N$ \\
  0           &    1           & $N-1$ \\
$\vdots$ & $\vdots$   & $\vdots$ \\
  0           &   $N$        & 0  \\ \hline
  1           &   0            & $N-1$ \\
$\vdots$ & $\vdots$   & $\vdots$ \\
  1           &   $N-1$    & 0 \\ \hline
  $N$      &    0           & 0
\end{tabular}
\end{center}
and the number of 3D HO basis states in the shell $N$ then reads
\begin{equation}
n_{\rm states}(N) = (N+1)+N+(N-1)+\cdots+1=\frac{1}{2}(N+1)(N+2) \; .
\end{equation}
Because parity is conserved, the basis can be separated into positive
and negative parity blocks. The dimension of the each block is
determined by summing up the number of states in even/odd-$N$ shells:
\begin{align}
n_{\rm pos.} &=\frac{1}{2} \sum_{k=0}^{k_{max}^+}{(2k+1)(2k+2)}
   = \frac{1}{6}\left(k_{max}^+ +1  \right)\left(k_{max}^+ +2 \right)\left(4k_{max}^+ +3  \right), \\
n_{\rm neg.} &=\frac{1}{2} \sum_{k=0}^{k_{max}^-}{(2k+2)(2k+3)}
 = \frac{1}{6}\left(k_{max}^- +1  \right)\left(k_{max}^- +2 \right)\left(4k_{max}^- +9  \right)\; ,
\end{align}
where $k_{max}^+ = [N_{max}/2]$ and
$k_{max}^- = [(N_{max}-1)/2]$, and the square brackets denote integer division.

\section{\label{App-pp}Transformation of the product of 1D
                   HO wave functions to the center-of-mass frame }

By multiplying the generating function for the Hermite polynomials
\begin{equation}
g(x,p,b) = e^{2xp/b-p^2} = \sum_{n=0}^\infty{\frac{1}{n!}p^n H_n(x/b)},
\end{equation}
with the factor
$\displaystyle \frac{1}{\sqrt{b}} \pi^{-1/4}e^{-x^2/2b^2}$, we obtain
the generating function for the HO wave functions:
\begin{equation}
\label{generating-HO}
\frac{1}{\sqrt{b}}\pi^{-1/4}e^{-x^2/2b^2+2xp/b-p^2} =
\sum_{n=0}^\infty{\eta_n(p)\phi_n(x,b) },
\end{equation}
where $\eta_n(p)$ denotes
\begin{equation}
\eta_n(p) = p^n \sqrt{\frac{2^n}{n!}} \; .
\end{equation}
For the product of two generating functions
\begin{equation}
\label{product-app}
g(x_1,p_1,b)g(x_2,p_2,b) =  \frac{1}{b}\pi^{-1/2}
e^{-\frac{1}{2b^2}\left(x_1^2+x_2^2\right) + \frac{2}{b}\left(x_1p_1+x_2p_2 \right)
     -\left(p_1^2+p_2^2\right) },
\end{equation}
a new set of coordinates is introduced:
\begin{equation}
\left. \begin{array}{c}
\tilde{x} = \frac{1}{\sqrt{2}} \left( x_1 - x_2\right)\\
\tilde{X}= \frac{1}{\sqrt{2}} \left( x_1 + x_2\right)
\end{array} \right\} \Longleftrightarrow
\left\{ \begin{array}{c}
x_1 = \frac{1}{\sqrt{2}} \left( \tilde{X} + \tilde{x}\right)\\
x_2 = \frac{1}{\sqrt{2}} \left( \tilde{X} - \tilde{x} \right)
\end{array}\right. ,
\end{equation}
with an analogous relation for the variables $p_1$ and $p_2$. The
exponent in Eq.~(\ref{product-app}) can now easily be expressed in terms
of the new coordinates:
\begin{equation}
g(x_1,p_1,b)g(x_2,p_2,b) = \frac{1}{b}\pi^{-1/2}
e^{-\frac{1}{2b^2}\left(\tilde{x}^2+\tilde{X}^2\right)
+ \frac{2}{b}\left(\tilde{x}\tilde{p}+\tilde{X}\tilde{P} \right)
     -\left(\tilde{p}^2+\tilde{P}^2\right) }=
g(\tilde{X},\tilde{P},b)g(\tilde{x},\tilde{p},b).
\end{equation}
By using the definition of the generating functions
\begin{equation}
\label{product2}
g(\tilde{X},\tilde{P},b)g(\tilde{x},\tilde{p},b)
=\sum_{N=0}^\infty{\eta_N(\tilde{P}) \phi_N(\tilde{X},b)}
  \sum_{n=0}^\infty{\eta_n(\tilde{p}) \phi_n(\tilde{x},b)} \; ,
\end{equation}
the coefficients $\eta_N(\tilde{P})$ and $\eta_n(\tilde{p})$ can be
expressed in terms of $p_1$ and $p_2$:
\begin{align}
\eta_N(\tilde{P}) &= \sqrt{\frac{2^N}{N!}} \tilde{P}^N =
\sqrt{\frac{1}{N!}} (p_2+p_1)^N
=  \sqrt{\frac{1}{N!}} \sum_{M=0}^N{ \left( \begin{array}{c} N \\ M \end{array}\right)
p_1^{N-M}p_2^M}, \\
\eta_n(\tilde{p}) &= \sqrt{\frac{2^n}{n!}} \tilde{p}^n =
\sqrt{\frac{1}{n!}} (p_1-p_2)^n
=  \sqrt{\frac{1}{n!}} \sum_{m=0}^n{ (-1)^{n+m}
\left( \begin{array}{c} n \\ m \end{array}\right)
p_1^{m}p_2^{n-m}}.
\end{align}
The product of generating functions Eq.~(\ref{product2}) then reads
\begin{equation}
\label{product3}
\sum_{n,N=0}^\infty{\phi_N(\tilde{X},b)\phi_n(\tilde{x},b)\sqrt{\frac{1}{N!n!}}
\sum_{M=0}^N\sum_{m=0}^n
{(-1)^{n+m}  \left( \begin{array}{c} N \\ M \end{array}\right)
    \left( \begin{array}{c} n \\ m \end{array}\right)p_1^{N-M+m}p_2^{M+n-m}}  }.
\end{equation}
With the introduction of  the auxiliary indices:
\begin{equation}
n_1 = N-M+m \quad \textnormal{and} \quad n_2=M+n-m,
\end{equation}
the product Eq.~(\ref{product3}) can be rewritten in the form:
\begin{align}
g(\tilde{X},\tilde{P},b)g(\tilde{x},\tilde{p},b)&=
\sum_{n_1,n_2=0}^\infty{p_1^{n_1}p_2^{n_2}}
\sum_{n,N=0}^\infty \phi_N(\tilde{X},b)\phi_n(\tilde{x},b)\sqrt{\frac{1}{N!n!}}
\times \nonumber \\
 &\times
\sum_{M=0}^N\sum_{m=0}^n
{(-1)^{n+m}  \left( \begin{array}{c} N \\ M \end{array}\right)
    \left( \begin{array}{c} n \\ m \end{array}\right)
    \delta_{n_1,N-M+m}\delta_{n_2,M+n-m}} .
\end{align}
One of the Kronecker symbols can be used to eliminate the sum over $M$:
\begin{align}
g(\tilde{X},\tilde{P},b)g(\tilde{x},\tilde{p},b) &= \sum_{n_1,n_2=0}^\infty{p_1^{n_1}p_2^{n_2}}
\sum_{n,N=0}^\infty
 \phi_N(\tilde{X},b)\phi_n(\tilde{x},b)\sqrt{\frac{1}{N!n!}}\delta_{n_1+n_2,N+n}
\times \nonumber \\
 &\times
\sum_{m=0}^n
{(-1)^{n+m}  \left( \begin{array}{c} N \\ N-n_1+m \end{array}\right)
    \left( \begin{array}{c} n \\ m \end{array}\right)  } .
\end{align}
A comparison with the equivalent relation for the product (cf.
Eq.~(\ref{product-app}))
\begin{equation}
g(x_1,p_1,b)g(x_2,p_2,b)=\sum_{n_1,n_2=0}^\infty{p_1^{n_1}p_2^{n_2}
\sqrt{\frac{2^{n_1+n_2}}{n_1!n_2!}} \phi_{n_1}(x_1,b)\phi_{n_2}(x_2,b) } \; ,
\end{equation}
leads to the final expression for the transformation of the
product of HO wave functions:
\begin{equation}
 \phi_{n_1}(x_1,b)\phi_{n_2}(x_2,b) = \sum_{N,n}{
 M_{n_1n_2}^{nN} \phi_N(\tilde{X},b)\phi_n(\tilde{x},b) } \; ,
\end{equation}
where $M_{n_1n_2}^{nN}$ are the coefficients of the transformation
\begin{equation}
M_{n_1n_2}^{nN} = \sqrt{\frac{n_1!n_2!}{n!N!}}\sqrt{ \frac{1}{2^{N+n}}}
\delta_{n_1+n_2,n+N}\sum_m{(-1)^{n+m}
           \left( \begin{array}{c} N \\ N-n_1+m \end{array}\right)
           \left( \begin{array}{c} n \\ m \end{array}\right)}\; .
\end{equation}
For the calculation of matrix elements of the pairing interaction,
the center-of-mass and relative coordinates are used
\begin{equation}
X \equiv \frac{1}{2}(x_1+x_2) = \frac{1}{\sqrt{2}}\tilde{X} \quad
\textnormal{and} \quad
x \equiv x_1-x_2 = \sqrt{2} \tilde{x} \; .
\end{equation}
The HO wave functions are expressed in terms of $X$ and $x$:
\begin{align}
\phi_N(\tilde{X},b) &= \phi_N(\sqrt{2}X,b) = \frac{1}{\sqrt{b}}
        \mathcal{N}_n H_n\left(\sqrt{2}X/b\right)e^{-2x^2/2b^2}
          = \frac{1}{\sqrt{2}} \phi_N(X,B) \; , \\
\phi_n(\tilde{x},b) &= \phi_n(x/\sqrt{2},b) = \frac{1}{\sqrt{b}}
        \mathcal{N}_n H_n\left(x/\sqrt{2}b\right)e^{-x^2/4b^2}
          = \sqrt{2}  \phi_n(x,b_r)\; ,
\end{align}
where we have defined the oscillator lengths $B=b/\sqrt{2}$ and
$b_r=\sqrt{2}b$. Finally, the product of two HO wave functions
expressed in terms of the center-of-mass and relative coordinates
reads:
\begin{equation}
\label{moshinsky}
 \phi_{n_1}(x_1,b)\phi_{n_2}(x_2,b) = \sum_{N,n}{
 M_{n_1n_2}^{nN} \phi_N(X,B)\phi_n(x,b_r) }.
\end{equation}

\clearpage
\clearpage
\clearpage
\clearpage
\clearpage
\clearpage


\end{document}